\documentclass[fleqn,usenatbib]{rasti}

\usepackage{newtxtext,newtxmath}

\usepackage[T1]{fontenc}

\DeclareRobustCommand{\VAN}[3]{#2}
\let\VANthebibliography\thebibliography
\def\thebibliography{\DeclareRobustCommand{\VAN}[3]{##3}\VANthebibliography}

\usepackage{graphicx}	
\usepackage{amsmath}	
\usepackage{bm}   

\usepackage[makeroom]{cancel} 
\usepackage{threeparttable} 
\usepackage{booktabs} 
\usepackage{subcaption}
\usepackage{algorithm} 
\usepackage{algpseudocode}
\usepackage{array} 

\title[t-FMPE for Time Series Data]{Fast Inference on Astronomical Time Series with Trans-Dimensional Flow Matching Posterior Estimation}

\author[N. van der Meulen et al.]{
Nina van der Meulen,$^{1}$\thanks{E-mail: ninavdmeulen1@gmail.com}
Tin Hadži Veljković,$^{2}$
Daniela Huppenkothen,$^{1}$
Benjamin Kurt Miller$^{2,3}$ and
\newauthor Christoph Weniger$^{3}$
\\
$^{1}$Anton Pannekoek Institute, University of Amsterdam, Science Park 904, 1098 XH Amsterdam, The Netherlands\\
$^{2}$Informatics Institute, University of Amsterdam, Science Park 900, 1098 XH Amsterdam, The Netherlands\\
$^{3}$GRAPPA Institute, Institute for Theoretical Physics Amsterdam, University of Amsterdam, Science Park 904, 1098 XH Amsterdam, The Netherlands}

\date{Accepted XXX. Received YYY; in original form ZZZ}

\pubyear{\the\year{}}

\begin{document}
\label{firstpage}
\pagerange{\pageref{firstpage}--\pageref{lastpage}}
\maketitle

\begin{abstract}
The analysis of time series plays an important part in the study of (fast) transient events, including gamma-ray bursts, magnetar bursts, fast radio bursts, and solar flares. A common approach is to decompose the time series into pulses and study the pulse characteristics, such as location and amplitude, in order to constrain physical models of the source and its environment. However, estimating both the number and characteristics of these pulses presents a trans-dimensional inference problem that traditional sampling methods such as Markov Chain Monte Carlo (MCMC) and Nested Sampling struggle to solve efficiently. Simulation-based inference methods, often incorporating machine learning techniques, provide an alternative approach when traditional approaches are insufficient. 
Here, we introduce \textit{trans-dimensional} Flow Matching Posterior Estimation (t-FMPE) implemented on a transformer architecture capable of efficient, amortized trans-dimensional inference on uniformly sampled univariate time series data. In this initial study, we apply the method to three test cases: simulated time series with known ground-truth parameters, observational data of Fast Radio Bursts and observations of X-ray bursts from magnetars. We show that t-FMPE achieves qualitative agreement with MCMC reference posteriors, successfully reproducing parameter correlations, as quantified through classifier two-sample tests. The trained network performs inference several orders of magnitude faster than MCMC and nested sampling, reaching sampling rates of 100 posterior samples per second for an 80-dimensional parameter space.
The results demonstrate potential of t-FMPE for large-scale analysis of time series datasets when traditional sampling methods become infeasible, and also enable inferring unbiased posteriors in the presence of observational biases such as dead time. 

\end{abstract}

\begin{keywords}
Machine Learning -- Fast Transients -- Simulation-Based Inference -- Transformers -- Time series -- Flow Matching
\end{keywords}



\section{Introduction}
High-energy astrophysical transients such as Fast Radio Bursts (FRBs), Gamma-Ray Bursts (GRBs), and X-ray bursts from magnetars encode a wealth of information about the underlying physics. In GRBs, the prompt emission is expected to carry information about the underlying engine driving the energy production and emission as well as internal jet dynamics \citep{rees1994, kumar2015}. In magnetars, X-ray bursts have been posited to be caused by magnetic reconnection or possibly neutron starquakes, potentially enabling studies of magnetic fields above the quantum-critical limit, and of the Equation of State of ultra-dense matter \citep[e.g.][]{thompson1995, lyutikov2003, perna2011, lander2016, kingma2017_adam_methodstochasticoptimization}. FRBs are currently the subject of intense study while their underlying driving engine and emission mechanisms remain unknown, though magnetars are a strong contender \citep[for recent reviews, see][]{petroff2022,zhang2023}. 

The light curves of these high-energy transients tend to be varied and often multi-peaked, indicating a complex, potentially stochastic underlying process \citep[e.g.~][]{bazzantini2024}. A wide array of models exist for these light curves, though most are empirical rather than strictly physically based. These models include stochastic processes \citep[e.g.][]{huebner2022}, often parametrized in the Fourier domain, as well as models that consider the observed light curve a superposition of smaller emission events. In particular in GRBs, the latter model has been used extensively for observational studies \citep{MEPSA-2015-guidorzi,MEPSA-user-1-2020,MEPSA-user-2-2019, MEPSA-user-3-2024,MEPSA-user-4-2025}, since it fits naturally with the expectation that collisions of shocks in the expanding jet produce a series of short, bright pulses, the superposition of which produces the observed light curve. For magnetar bursts, one hypothesis considers the observed time evolution to be the result of a cascade of reconnection events \citep{Huppenkothen_2015}. For both, the precise emission mechanism is largely unknown, and thus studies default to simple empirical shapes (e.g. fast-rise, exponential decay profiles or Gaussians) to model the individual components, with some (existing, but limited) physical interpretation of these shapes. In FRBs, individual pulses are often modelled using Gaussians \citep{fitburst_algorithm_Fonseca_2024}, though modified by physical effects such as scattering of radio photons in the intervening medium. Additionally, the ``Sad Trombone'' effect has been observed, a systematic downward drift in sub-pulse frequency over time, which has been identified as a critical diagnostic for the emission mechanism \citep{sad_trombone_Hessels_2019}.

Independent of the precise shape of each pulse in the light curve, distributions of the pulse parameters are relevant to the physics and therefore important to infer. These include waiting time distributions between pulses, distributions of the rise time and skewness of the pulses, and the distribution of amplitudes. For example, in GRBs, the distributions of pulse widths and the intervals between them have been shown to follow log-normal distributions, which in turn has been linked to a stochastic process governed by internal engine variability \citep{norris1996,bhat2012}.
In FRBs, the power law-like shapes of the energy distribution have often been linked to Self-Organized Criticality, while the detection of a break in the energy distribution of repeating FRBs serves as a critical diagnostic for the emission site \citep{wu2025}. In order to infer these distributions of parameters, we must know how many pulses are present in the light curve, and then infer their parameters. Most existing approaches separate these steps in some way. For example, many studies rely on visual identification of peaks before subsequent fitting of a component model, however, this approach relies on a precise understanding of the uncertainties of the data, and lacks statistical robustness. A simple peak algorithm such as \texttt{find\_peaks} implemented in SciPy \citep{scipy2020} identifies pulses by comparing each bin to its immediate neighbours, defining a peak as a value greater than its left and right neighbours. This approach, too, comes with statistical robustness concerns, particularly for noisy data. Constraints such as minimum width, amplitude or prominence can be applied to avoid fitting peaks to noise, but the result remains a single point estimate of the number of peaks, without any measure of uncertainty. Additionally, it may require substantial manual tuning. For example, \citet{curtin2025morphology_35_frbs_repeaters} report that when the output of \texttt{find\_peaks} did not match the number of peaks identified by eye, the results were manually adjusted. Such practices should ideally be avoided, as they introduce significant human bias and provide no framework for quantifying uncertainty.

The Multiple Excess Peak Search Algorithm (MEPSA) \citep{MEPSA-2015-guidorzi} was originally developed for GRBs, and searches for peaks by testing whether segments of a time series conform to any of the 39 pre-defined peak patterns. This is evaluated across different time resolutions of the same profile, allowing it to detect peaks on a variety of time scales. Some of the variables provided in MEPSA’s output include the location (peak time), timescale and SNR of each peak. This method has been shown to have a significantly lower false positive rate than naive peak finding algorithms and has since been widely used in the analysis of GRB signals, and a limited number of FRBs \citep{MEPSA-user-1-2020, MEPSA-user-2-2019, MEPSA-user-3-2024, MEPSA-user-4-2025}. MEPSA has a few limitations: it assumes uniformly sampled time series with Gaussian noise, where background noise or trends have already been removed. The 39 pre-defined patterns are specifically tuned to patterns observed in GRB data and might require adjusting when applying the algorithm to other transients. For example, \citet{hewitt_2023_dense_forests_microshots_bursts} present a set of three particularly complex FRBs with dozens of peaks per burst, and suggest that there may be two different types of pulses in the data. Additionally, the algorithm does not naturally model correlations between pulse parameters, which in turn may lead to biases in the inferred parameters, especially when pulses overlap. An updated version, FAST-MEPSA, has recently been published \citep{FAST_MEPSA_MAISTRELLO2026101040}.

A general challenge with algorithms that first attempt to find the peaks, and then fit the resulting model to the data, is that these approaches make it challenging to robustly propagate uncertainties in the peak detection into the inference of the parameters. In GRBs, \citet{baldeschi2015} showed that the lognormality of the inferred distribution of interpulse intervals could be an artefact of the low efficiency of peak search algorithms at small intervals. The uncertainty in peak detections stems from the statistical noise inherent in our data: often, it is challenging for both humans and algorithms to determine whether a feature in the data is a real peak or a pattern produced by random chance. To address this issue \citet{magnetron-software} built a model for X-ray bursts from magnetars that inferred both the number of pulses in the light curve along with the parameters of each pulse using Bayesian hierarchical modeling and \textit{trans-dimensional nested sampling}: in each sampling step, components to model pulses could be added or subtracted, and thus the dimensionality of the parameter space can be changed dynamically. The final posterior distribution is a joint distribution over both the number of pulses and the pulse parameters, accurately propagating our uncertainty in the number of peaks into the inference of the parameters. While powerful, this approach proved to be computationally expensive, and thus cannot be applied to the rapidly growing data sets observed e.g.~for FRBs. 

In this paper, we present a new approach to this problem using \textit{simulation-based inference} (SBI). In this approach, the posterior distribution is approximated from pairs of parameters and simulated data, and the complex, high-dimensional distribution is approximated using neural networks, here in the form of \textit{flow matching}. We take the core idea from \citet{Huppenkothen_2015}, and enhance it through neural network-based sampling in order to improve computational speed and efficiency. The model is \textit{amortized}: once trained, generating $\sim$1000 samples for a single burst takes of the order of seconds, with the exact runtime depending on the number of peaks and the network settings. As a consequence, generating distributions for large samples of bursts becomes feasible. In addition, because this framework relies on simulators, rather than a known likelihood, it is possible to include selection effects and instrumental biases such as \textit{dead time}, which distort the data in ways that are difficult to include in an analytic likelihood. 

The paper is structured as follows: The next section describes the observational radio and X-ray data relevant to this study. In section \ref{sec:methods} we discuss the generation of artificial light curves and introduce the proposed trans-dimensional Flow Matching Posterior Estimation (t-FMPE) method in detail, including a brief introduction to flow matching, a description of the network architecture, and the training and inference procedures. In Sections \ref{sec: results-sim-data} and \ref{sec:OD-results}, we apply t-FMPE to simulated data, observations of X-ray bursts and FRB profiles: We compare the resulting posteriors with those obtained through Markov Chain Monte Carlo (MCMC) sampling in fixed-component settings (number of components $N$ fixed) and provide a direct comparison with the trans-dimensional nested sampling (t-NS) approach in \cite{Huppenkothen_2015} for the trans-dimensional case. Finally, we apply t-FMPE to observational FRB profiles and find good agreement between the data and posterior samples.

Overall, we find that by implementing this novel SBI method, t-FMPE, on a transformer architecture, we are able to quickly infer posterior distributions over the number of peaks and peak parameters for a given light curve. For several test cases, the generated posterior samples show good qualitative agreement with traditional sampling methods, while requiring only $\sim$1\% of the computational cost. Future challenges include global validation of the neural posterior to evaluate the model performance over a large set of light curves, for example through global classifier two-sample tests on joint samples or coverage tests such as simulation-based calibration \cite{lopez-paz-2018-revisiting-c2st, talts-2020-sbc-validatingbayesianinferencealgorithms}.  Additional future work includes incorporating more complex, physically motivated models to enable thorough testing of emission and source hypotheses of fast transients, as well as extending the approach to time series of variable length and two-dimensional time series that record intensity over several frequency channels.

\section{Observational Data}
\label{sec:OD}

All light curves in this study are represented by a discretized time series with $K=1000$ evenly spaced time bins giving a time series $\bm y=\{y_k\}_{k=1}^K$, where $y_k$ is the value in bin $k$. This time series may contain a simulated measurement, or observational flux (radio) or count (X-ray) measurements. The noise-free model of a light curve is represented by an array of $K$ modelled rates, $\bm \lambda =\{\lambda_k\}_{k=1}^K$. For this pilot study, we chose 1000 time bins for simplicity, and will address the limitations that choice imposes in Section \ref{sec:obsprocessing}.

\subsection{Magnetar Burst}
We use X-ray data from one of the magnetar bursts analysed by \citet{Huppenkothen_2015} to enable a comparison with their trans-dimensional nested sampling algorithm known as \texttt{magnetron}. This burst was observed by the Gamma-Ray Burst Monitor (GBM) onboard the \textit{Fermi} Space Telescope \citep{meegan2009} and originates from the Galactic Soft Gamma Repeater (SGR) J1550-5418 \citep{camilo2007,kaneko2010}. Magnetars are known emit short bursts of X-ray emission that, like FRBs, exhibit a high degree of variability and complex temporal structure. Such bursts are known to occur during periods of high activity, referred to as burst episodes or storms. 

The X-ray emission is recorded as events representing the arrival of individual photons as a function of time. The Fermi/GBM instrument contains 12 NaI detectors sensitive to photons energies between 8 keV and 4 MeV. In this dataset, the photon energy was limited to 8-200 keV, where magnetars have been observed to be the most active. 

The selected burst from SGR J1550-5418 was binned at a time resolution of 0.5 ms with 1664 time steps covering a time period of 832 ms. Before applying t-FMPE, the observed counts were down-sampled by a factor of two and symmetrically padded with 84 bins of Poisson noise ($\lambda_{bkg}=3$) on either side to get $K=1000$ bins in total, matching the number of bins in the simulated light curves used in training.

\subsection{Radio Data}
We perform inference on three multi-peaked FRBs: FRB20190115B, FRB20190122C and FRB20190124F. These FRBs were detected by the Canadian Hydrogen Intensity Mapping Experiment (CHIME; \citealt{chimepathfinder,chimefrb}) in 2019 and are named after their detection date (YYYYMMDD) with a letter appended to distinguish bursts detected on the same day. These bursts were selected to include a range of temporal complexity to test t-FMPE across several FRB profiles from the CHIME/FRB baseband catalog \citep{frbcollaboration2024updatingchimefrbcatalogfast}. 

CHIME is a radio interferometer consisting of four large cylindrical dishes \citep{2018_CHIME_telescope}. CHIME periodically publishes a burst catalogue and maintains a real-time detection system for FRB candidates. Their first \textit{baseband} catalogue supplies channelized raw voltage (baseband) data\footnote{See \url{https://www.chime-frb.ca/baseband-catalog-1} .} of 140 FRBs from Catalog 1 \citep{first_chime_catalog_2021, frbcollaboration2024updatingchimefrbcatalogfast}, enabling studies of dynamic spectra\footnote{A two-dimensional histogram of signal intensity over time and frequency} with high time resolution and high sensitivity. The catalogue also includes several estimated quantities in tabular format, such as the burst's inferred Dispersion Measure (DM), number of components (manually estimated), and burst duration. This information is summarized for the selected FRBs in table \ref{tab:frb-data}. 

The baseband data contains radio signals recorded between 400 and 800 MHz over 1024 frequency channels at a time resolution of 2.56 $\mu$s. Unlike X-ray data, which is recorded in photon counts, radio data is recorded as intensity or flux, and the flux uncertainties are assumed to be Gaussian. 

\begin{table}
    \centering
\caption{FRB sample used in this study. Tabular data adapted from CHIME's baseband catalog. $N$ is the number of distinct peaks in the profile.}
\label{tab:frb-data}
    \begin{tabular}{ccccc}\toprule
         Name&  $N$& \shortstack{Duration \\ (ms)}&\shortstack{DM \\ (pc cm$^{-3}$)} & {\shortstack{Down \\ factor}}\\\midrule
         FRB20190115B&  3& 5.90 &748 &8\\
         FRB20190122C&  7& 27.75 &690 &16\\
         FRB20190124F&  6& 2.79 &255 &2\\\bottomrule
    \end{tabular}
\end{table}

The raw voltage data is pre-processed into one-dimensional light curves that match the training data in time resolution and number of bins. The dynamic spectrum is first de-dispersed using the DM provided in the catalogue, to correct for dispersion effects caused by the intervening medium, and then downsampled by a factor shown in table \ref{tab:frb-data}. 

Next, frequency channels that are dominated by radio frequency interference (RFI) are removed. This is an iterative process whereby all frequency channels with standard deviations over five times the average are removed. This is done ten consecutive times, where removed frequency channels are filled with \texttt{NaN}s. 

For each burst, the recorded time interval often extends beyond the burst's actual duration, as $\sim100$ms of baseband data is saved for all candidate FRBs \citep{frbcollaboration2024updatingchimefrbcatalogfast}. Therefore, data is cut such that the burst is approximately at the centre of the time window and the total number of bins corresponds to the training data ($K=1000$). 

The flux values are converted to Janksy units and summed over the frequency dimension to convert the two-dimensional dynamic spectrum into a one-dimensional time series. Finally, the light curve is normalized such that the background noise is approximately Gaussian, i.e. $y_k = \lambda_k + \epsilon$ with $\epsilon \sim N(0, 1)$. 

\section{Methods}\label{sec:methods}
\subsection{The Burst Model and Training Data} \label{sec:component-model-intro}

\begin{figure}
    \centering
    \includegraphics[width=0.7\linewidth]{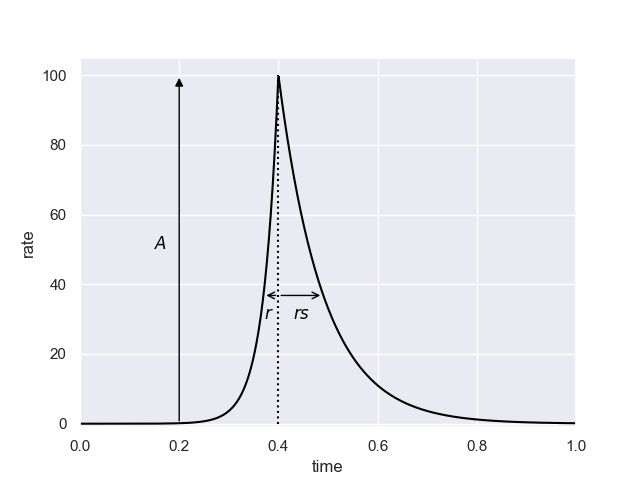}
    \caption{Single noise-free burst component following equation \ref{eqn:burst-component}. The component has amplitude $A$, rise time $r$ and a fall time given by the product of the rise time and the skewness $s$. The peak-time $t_0$ is denoted by the dotted line. Figure adapted from \citet{Huppenkothen_2015}.}
    \label{fig:single-component-noise-free}
\end{figure}
Light curves of fast transients with complex temporal variability are difficult to capture with a single functional form. Therefore, a more general way to model them is by using a component model \citep{Huppenkothen_2015}. This model deconstructs bursts into $N$ simple shapes, parametrized by some function $\lambda_n(t)$. There are many empirical choices for the functional form of $\lambda_n(t)$. For FRBs for example, a skewed Gaussian is commonly used. In this study, the selected shape is a double-sided exponential peak. This is motivated by its simplicity, its use in prior work (allowing for direct comparison of results) and its demonstrated success in modelling magnetar X-ray bursts \citep{Huppenkothen_2015}. A single component $n$ is defined as

\begin{equation}
    \lambda_n(t)=\begin{cases}
      Ae^{\frac{t-t_0}{r}} & \text{if $t \leq t_0$}\\
      Ae^{-\frac{t-t_0}{rs}} & \text{if $t > t_0$}
    \end{cases}\;\;,
    \label{eqn:burst-component}
\end{equation}
with four free parameters: amplitude $A$, peak time $t_0$, rise time $r$, and skewness parameter $s$, shown in figure \ref{fig:single-component-noise-free}. 

To get a full noise-free light curve $\lambda(t)$, $N$ components are added together and a constant background rate $\lambda_{\rm bkg}$ is added to account for the baseline level flux in observations

\begin{equation}
    \lambda(t) = \lambda_{\rm bkg} + \sum^N_{n=1} \lambda_n(t)\;.
    \label{eq:noise-free-curve}
\end{equation}

\begin{figure}
    \centering
    \includegraphics[width=\linewidth]{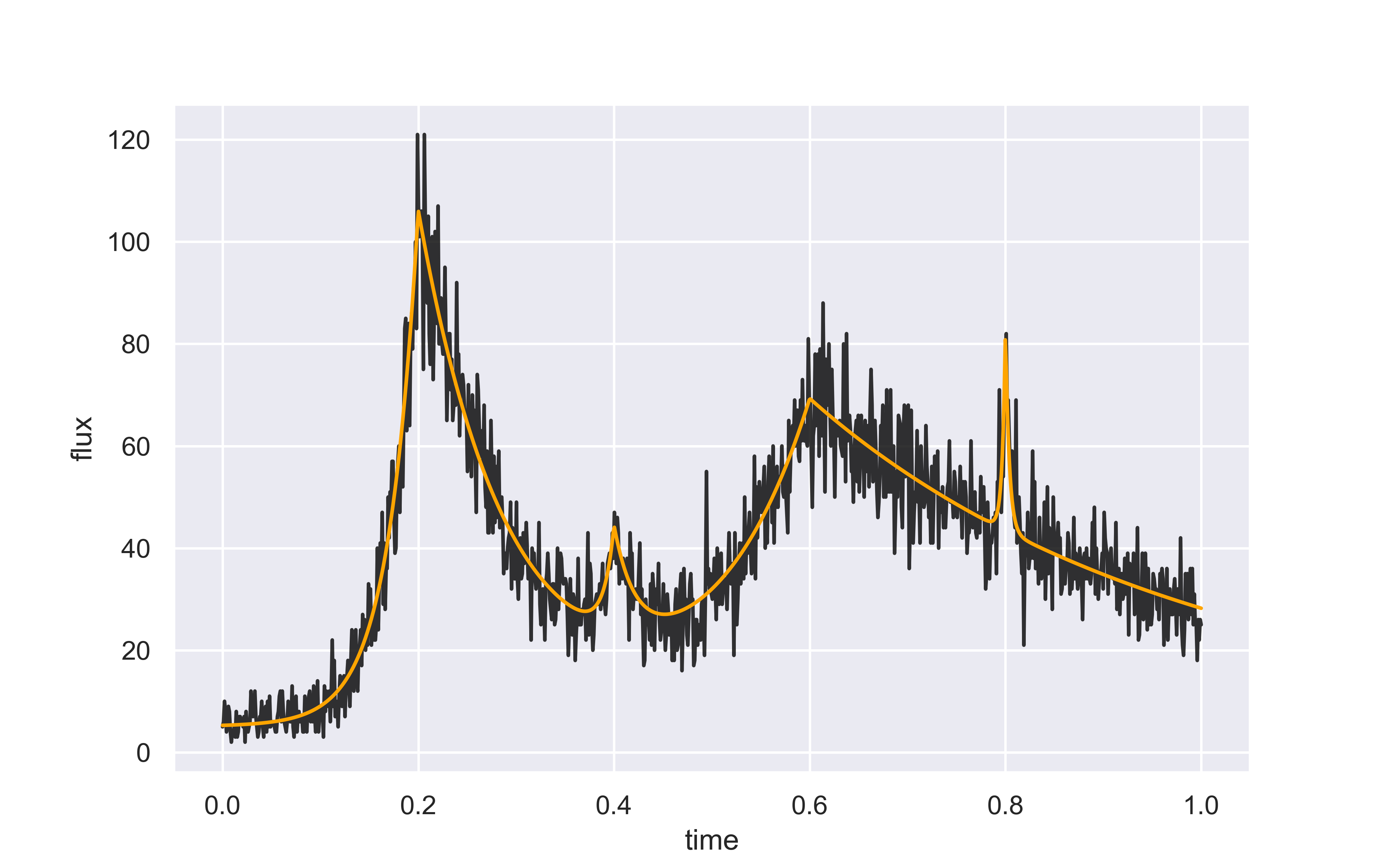}
    \caption{Example burst sampled from the simulator consisting of $N=4$ components. The ground-truth noise-free flux is shown in orange.}
    \label{fig:example-burst-four-components}
\end{figure}

Figure \ref{fig:example-burst-four-components} shows an example burst that was generated with this model with Poisson noise added. From the simulated (and eventually, observational) data, we seek to infer the key parameters of each component, as well as the total number of components. This enables a quantitative analysis of temporal variability, which can hold important clues needed to further constrain emission and source models, as discussed in \citet{what_frb_variability_tells_us}. 

A fundamental challenge of this model is the fact that the number of parameters to be inferred is not known in advance, as it depends on the (unknown) number of burst components. As a result, the parameter space is trans-dimensional and generally high-dimensional, making standard inference methods difficult to apply. 

\begin{table}
\centering
\begin{threeparttable}
\caption{Variable symbols and meaning.}
\label{tab:variables}
\begin{tabular}{r|p{3.5cm}cl}
\toprule
 Symbol &Description & Dimension&Used in\\ 
 \midrule
 $\bm y$, $\{y_k\}$&Light curve; time series of brightness measurements& $K(=1000)$ &FM, MCMC\\[15pt]
 $\bm \lambda$, $\{\lambda_k\}$&Noise-free light curve& $K(=1000)$ &FM, MCMC\\[5pt]
 $\lambda_{\rm bkg}$& Background rate& scalar &FM, MCMC\\[5pt]
 $t_0$ & Peak position & scalar & FM, MCMC \\ [5pt]
 $A$ & Amplitude & scalar & FM, MCMC \\ [5pt]
 $r$ & Rise time & scalar & FM, MCMC \\ [5pt]
 $s$ & Skewness & scalar & FM, MCMC \\ [5pt]
 $\bm \theta$, $\{\bm \theta_n\}$ & Set of parameter vectors& $4 \times N_{\rm max}$&FM, MCMC\\[5pt]
 $N_{\rm max}$&Maximum number of burst components in training data& scalar&FM\\[15pt]
 $N_{\rm true}$& True number of burst components& scalar&FM, MCMC\\[5pt]
 $\tau$ & flow matching time& scalar&FM\\[5pt]
 $\bm \theta_n$& Burst parameters of component $n$ & $4$&FM\\[5pt]
 $\bm  \theta_\tau$& Interpolated state between $\bm \theta_0$ and $\bm \theta_1$ at time $\tau\in [0,1]$& $4 \times N_{\rm max}$&FM\\[15pt]
 $\bm \theta_0$& Initial sample from the base distribution $\mathcal{N}(0, I_n)$ & $4 \times N_{\rm max}$&FM\\[15pt]
 $\bm \theta_1$& Final state at $\tau=1$& $4 \times N_{\rm max}$&FM\\[5pt]
 $\bm \theta_e$& Encoded parameter vector& $d_\theta$&FM\\[5pt]
 $\bm \tau_e$& Encoded flow matching time& $d_\tau$&FM\\[5pt]
 $\bm y_e$& Encoded time series& $d_y$&FM\\[5pt]
 $\bm u_\tau^\phi$& Parameterized vector field& $4 \times N_{\rm max}$&FM\\[5pt]
 $p_\tau$ & Probability path& -&FM\\[5pt]
 $\sigma_{\rm min}$ & Minimum variance of Gaussian path at $\tau=1$ & scalar&FM\\[5pt]
 $T_i$& Token at position $i$ in the sequence& $d_t$&FM\\[5pt]
 $\phi$& Network parameters& - &FM\\[5pt]
\bottomrule
 
\end{tabular}

\end{threeparttable}

\end{table}

\subsubsection{Notation}
For reference, an overview of the variables introduced in the following sections is provided in table \ref{tab:variables}. We summarize the parameters to be inferred in an ordered set, $\bm \theta =\{\bm \theta_n\}$, containing sub-vectors
\begin{equation}
    \bm{\theta}_n = \begin{bmatrix}
           t_0^n\:\; 
           A_n \;\;
           r_n \;\;
           s_n
         \end{bmatrix}\;,
\end{equation}
where $n$ indicates the index of the burst component, and for component $n$, the corresponding parameters are the peak time $t_0^n$, amplitude $A_n$, rise time $r_n$ and skewness $s_n$, illustrated in figure \ref{fig:single-component-noise-free}. The set contains $N$ parameter vectors, where $N$ is the number of burst components. We require the parameter vectors in the set to be chronologically ordered ($t_0^{n+1} > t_0^n$) in order to avoid a common problem where any component could model any feature in the data, leading to highly multi-modal posteriors. The skewness $s$ is defined as the ratio between the rise time $r$ and fall time $f$, $s = f / r$, such that a skewness larger than one indicates that the fall time is longer than the rise time.

\subsubsection{Adding Noise}\label{sec:adding noise}
The added noise depends on the kind of fast transient that is modelled. To simulate X-ray measurements, the noise-free light curve $\{\lambda_k\}$ is perturbed with Poisson noise by drawing a measurement from a Poisson distribution for each bin
\begin{equation}
    y_k \sim \text{Poisson}(\lambda_k)=\frac{\lambda_k^{x}e^{-\lambda_k}}{x!}\;.
    \label{eq: poisson-sampling}
\end{equation} 
For observational radio data the noise is normalized to be approximately Gaussian, hence the measurements are simulated by adding Gaussian noise to the modelled light curve
\begin{equation}
    y_k = \lambda_k +\epsilon \;\;\;\;\text{where} \;\;\epsilon\sim \mathcal{N}(0, 1)\;.
    \label{eq:gaussian-noise}
\end{equation} 

\subsubsection{Training Data Generation}
\label{sec: training-data}
To train the network, many labelled light curves $\bm\theta, N, \bm y \sim p(\bm\theta)p(N)p(\bm y\mid \bm\theta, N)$ are generated. As light curves can be generated quite efficiently with the component model, batches of training data are generated on-the-fly during training. This way, there is no concern of overfitting and the training dataset can be as large as necessary to produce reliable posteriors. 

Simulated light curves are generated in batches, $Y \in {\rm I\!R}^{B\times K}$, with $B=$ batch size and $K=1000$, the number of bins in the time series. A maximum number of components $N_{\rm max}$ is specified beforehand, which affects the size of the parameter set $\bm\theta \in {\rm I\!R}^{4\times N_{\rm max}}$. Contributions of each component to the total flux, given by equation \ref{eqn:burst-component}, are computed separately and iteratively summed. The number of burst components $N$ in the training data varies. This is done by sampling the number of components from discrete prior distribution $U(1, N_{\rm max})$ and setting the amplitude of non-contributing components to zero ${A_{n > N}} \gets 0$.  After summing all contributions, a fixed background rate $\lambda_{\rm bkg}$ is added, which changes between experiments further described in Section \ref{sec: results-sim-data} between 0, 3 or 5. Finally, the simulated flux is perturbed with noise, which is configured to be either Poissonian (X-ray; equation \ref{eq: poisson-sampling}) Gaussian (radio; equation \ref{eq:gaussian-noise}).

\subsection{Priors}
Each burst parameter is assigned a prior distribution, summarized in table \ref{tab:priors}. In t-FMPE they are used during training to draw simulated time series. The prior for the number of components is only relevant for trans-dimensional settings (t-FMPE; t-NS). This component prior is a discrete uniform distribution with an the upper limit $N_{\rm max}$. 

To account for the fact that the rise time and amplitude can span multiple orders of magnitude, these parameters are sampled from log-uniform distributions, such that each order of magnitude is represented equally. Instead of sampling $A$ and $r$ directly , this is done by sampling $\log_{10} (A)$ and $\log_{10}{(r)}$ from uniform distributions, and exponentiating to recover $A$ and $r$ for the burst model.

Peak times are sampled from an ordered distribution by first drawing $\{t_0^n\}_{n=1}^{N} \sim U(t_{\rm min}, t_{\rm max})$ and then sorting those values. This ordering addresses the common label switching problem and thus removes multi-modalities and degeneracies from the marginal posteriors of $\{t_0^n\}$, which can otherwise complicate MCMC convergence. The peak time range is set to $[0.2, 0.8]$ to avoid having to infer parameter for components that lie mostly outside of the considered time frame, $0 \leq t \leq 1$.

These priors were selected to generate light curves that reasonably resemble observational data and are defined with strict bounds. In the context of practical applications outside of this pilot work, the choice of prior might require more careful consideration.
For instance, the skewness is constrained to be $s\geq 1$ meaning that the fall time of a component, defined as $rs$ (see figure \ref{fig:single-component-noise-free}), is assumed to outlast the rise time. This aligns with expectations based on scattering effects in radio data, but this prior could be extended to accommodate intrinsic emission mechanisms where the rise time may exceed the fall time.

\begin{table}
\centering
\begin{threeparttable}
\caption{Prior distributions of burst parameters.}
\label{tab:priors}
\begin{tabular}{m{6.5em}cll}
\toprule
Description & Symbol & Prior & Range\\ 
\midrule
Number of burst components & $N$ & DiscreteUniform & $[1,N_{\rm max}]$ \\[7pt] 
Peak position of component $n$ & $t_0^n$ & Uniform$_{(n)}$\tnote{*} & $[0.2, 0.8]$ \\[7pt]
Rise time & $r$ & LogUniform & $[3\hspace{-0.15em} \cdot \hspace{-0.15em}10^{-3}, 0.1]$  \\[2pt]
Amplitude & $A$ & LogUniform & $[1, 300]$ \\[2pt]
Skewness & $s$ & Uniform & $[1, 6]$ \\[2pt]
\bottomrule
\end{tabular}

\begin{tablenotes}
\item[*] Uniform$_{(n)}$ represents the probability density function of the $n$-th order statistic of Uniform$(0.2, 0.8)$. 
\end{tablenotes}
\end{threeparttable}

\end{table}

\subsection{MCMC Posterior Estimation}\label{sec:mcmc-method}
As trans-dimensional MCMC sampling is non-trivial to implement and computationally demanding, we implement a simple MCMC sampler to compare parameter distributions for light curves with a fixed, known number of components. Therefore, MCMC sampling is performed primarily on simulated data where $N$ is known a priori in order to verify if t-FMPE is working correctly before moving to trans-dimensional settings. For its implementation, we use \texttt{emcee}'s \texttt{EnsembleSampler}, which is widely used in astrophysical publications \citep{emceepaper}.  

\subsubsection{Posterior distribution}
Since all the priors are uniform, the log posterior is simply equivalent to the log likelihood, except when walkers move out of bounds or violate the peak time ordering. In those cases the log posterior is set to $-\infty$.

Depending on the kind of noise in the time series, the likelihood is either Poisson or Gaussian.
The Poisson likelihood is defined as 

\begin{equation}
   p(\bm y\mid\bm \theta) =  \prod_{k=1}^K \frac{e^{-\lambda_k}\lambda_k^{y_k}}{y_k!}\;,
\end{equation}
with $y_k$ the simulated counts in bin $k$ and $\lambda_k$ the true rate in bin $k$. 

For the FRBs, the likelihood is given by the Gaussian likelihood 
\begin{equation}
   p(\bm y\mid \bm \theta) =  \prod_{k=1}^K \frac{1}{\sqrt{2\pi}\sigma}\exp\left[-\frac{(\lambda_k - y_k)^2}{2\sigma^2}\right]\;,
\end{equation}
where $\sigma = 1$ such that the log-likelihood simplifies to
\begin{align}
    \mathcal{L}_{G} 
    &= -\sum_{k=1}^K (\lambda_k - y_k)^2\;,
    \label{eq:gauss-log-likelihood}
\end{align}
where the term independent of model parameters has been neglected once again. 

The first part of the sampling algorithm is dedicated to a burn-in phase.  Initial walker positions are sampled from the prior and the \texttt{EnsembleSampler} is run for a tunable number of steps. After this preliminary run, walkers with relatively low likelihood (lower than the highest log probability so far, rounded down to the nearest multiple of 10) are repositioned to areas with the highest encountered likelihood, with a small amount of added noise
\begin{equation}
    x_{\rm unlikely} \gets x_{\rm best} \cdot \epsilon \;\;\;\;\; \text{with } \epsilon \sim U(0.9, 1.1)\;. 
\end{equation}
This process helps to prevent walkers from getting trapped in low-probability regions. After repositioning, the walkers are run for 500 burn-in steps. All samples from this phase are discarded, and the final chain is initialized from the resulting starting position. 

Sampling is continued until either a maximum number of iterations is reached, or until the chain length exceeds 100 times the estimated integrated autocorrelation time for all chains, ensuring a sufficient amount of uncorrelated samples. 

\subsection{Simulation-Based Inference}
The proposed t-FMPE method is an example of simulation-based inference (SBI). SBI refers to methods that use simulated data to approximate the posterior, without needing a tractable likelihood. These methods have gained attention in recent years, since formulating an accurate likelihood function for the data $y$ is often not trivial \citep{frontier_of_SBI_Cranmer_2020}. Most modern SBI techniques incorporate machine learning in some way, as they address several limitations of classical SBI techniques like Approximate Bayesian Computation (ABC), including poor scalability to high-dimensional problems and the lack of amortization \citep{frontier_of_SBI_Cranmer_2020}. An example strategy is to train neural density estimators such as discrete normalizing flows, which can transform simple base distributions into complex posteriors through a series of invertible transformations \citep{papamakarios2021normalizingflowsprobabilisticmodeling}. One of the advantages of these neural methods is that many allow for \textit{amortized inference}: Once trained, the computational inference cost is low, unlike ABC and traditional sampling methods that require many simulations for each new observation. There are still some limitations however: discrete normalizing flows require invertible transformations and computation of Jacobians, restricting the type of architecture that can be used, and continuous normalizing flows involve expensive loss evaluation.

These issues are solved in flow matching posterior estimation, a recent and promising development in this field. FMPE carries improved scalability and flexibility compared to normalizing flows, as the loss can be estimated efficiently and there are no restrictions on architecture at all \citep{dax2023flowmatchingscalablesimulationbased}. Consequently, flow matching is well suited for trans-dimensional inference problems, where expressive network architectures are required. 

\begin{figure}
    \centering
    \includegraphics[width=0.7\linewidth]{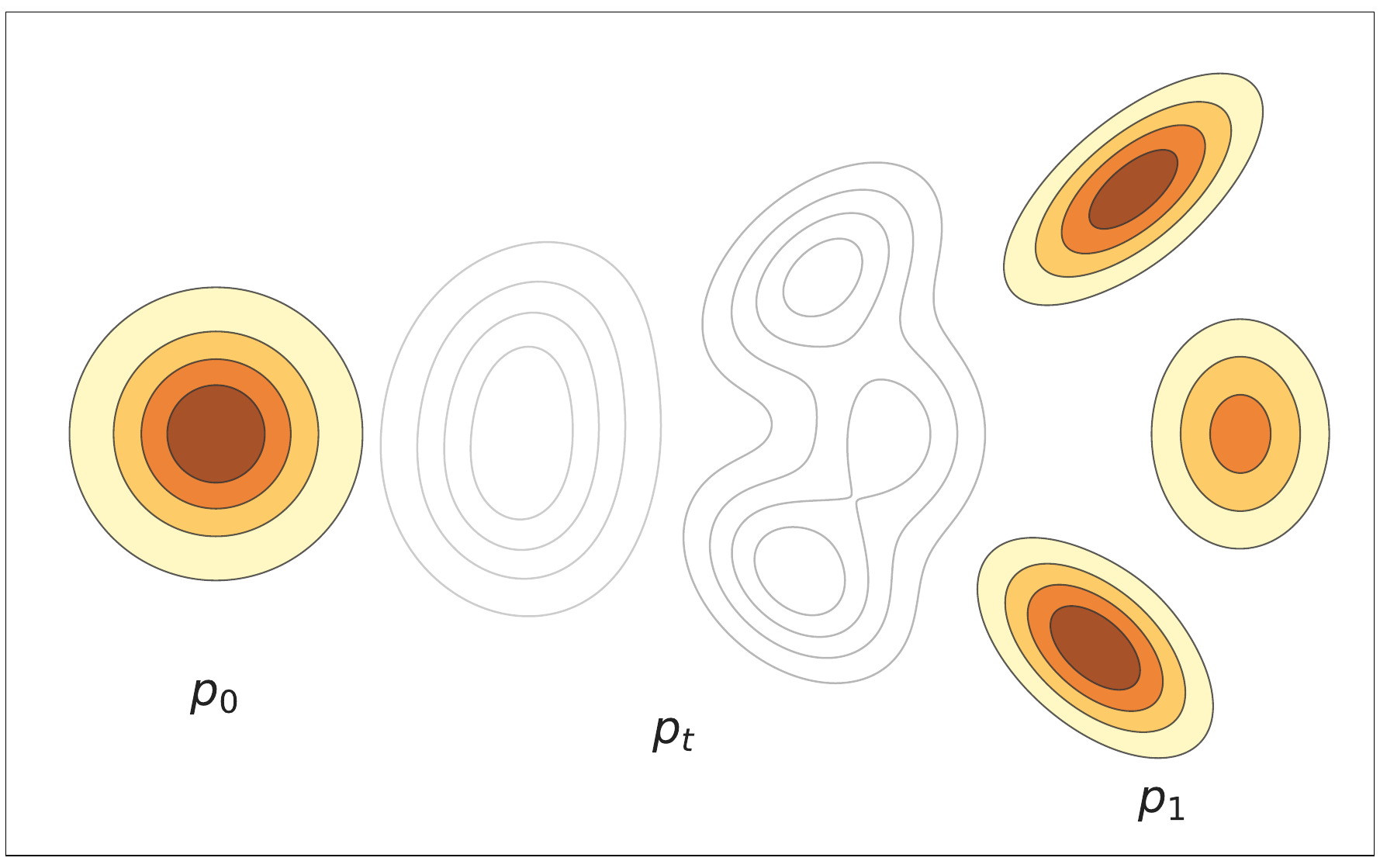}
    \caption{Flow matching concept. An initial distribution $p_0$ is continuously transformed into distribution $p_1$ via a time-dependent vector field, where $p_1$ may be any distribution of interest.}
    \label{fig:flow-matching-concept}
\end{figure}

\subsection{Flow Matching}\label{sec:FM-detailed-explanation}
Generally, flow matching is used to learn a time-dependent vector field $\bm u^\phi_\tau(\bm\theta)$\footnote{To avoid confusion between the time variable $t$ in context of time series, and time in flow matching ODEs, time is denoted by $\tau$ in the context of flow matching.}  that is capable of transforming samples from a trivial initial distribution $\bm\theta_0\sim p_{\rm init}(\bm\theta)$, into samples from a more complex target distribution $\bm\theta_1 \sim p_{\rm target}(\bm\theta)$, as illustrated in figure \ref{fig:flow-matching-concept}. Here, we provide a short overview of flow matching and how it can be adapted for posterior estimation, and refer to \citet{MIT_lecture_notes_FM_diffusion} for a pedagogical introduction to flow matching. 

Central to flow matching is the following ordinary differential equation (ODE)
\begin{equation}
    \partial_\tau \Theta_\tau = u_\tau(\Theta_\tau)\;,
\end{equation}
where $\Theta$ represents a trajectory in time, and $\Theta_\tau$ is the state of this trajectory at time $\tau$. The solution of this ODE is called the flow $\psi$  
\begin{equation}
\Theta_\tau = \psi_\tau(\bm\theta_0) \;\;\;\; \text{with   }\;\; \bm\theta_0 \sim p_{\rm init}(\bm\theta)\;.     
\end{equation}
In most cases, explicitly finding a closed-form expression of the flow is not possible. However, for a given vector field, the solution can be approximated via numerical integration methods. In flow matching, this vector field is parametrized and optimized to induce a flow from $p_{\rm init}$ to an arbitrary target distribution $p_{\rm target}$. In the case of FMPE, the target distribution is the posterior $p(\bm\theta\mid \bm y)$. The training objective focuses on minimizing the mean squared difference between the target and parametrized vector field for a single time step \citep{lipman2023_flowmatching}. This makes it more efficient than continuous normalizing flows, where loss evaluation demands integration of the vector field over all time steps.  

To sample from the trained flow matching model, we sample an initial state $\theta_0$ from the base distribution, and use the parametrized vector field to numerically integrate this state up to $\tau=1$ with a forward Euler scheme
\begin{equation}
   \bm \theta_{\tau+\Delta \tau} = \bm\theta_\tau + \bm u^\phi_{\tau}(\bm\theta_\tau) \Delta \tau \;,
    \label{eq:Euler_FM}
\end{equation}
where $\bm u^\phi_\tau$ is the vector field parametrized by network parameters $\phi$. This means inference requires many network evaluations rather than a single forward pass. The number of evaluations, and thus the computational cost of inference, scales linearly with the number of integration steps, which is typically chosen empirically to balance sample accuracy and efficiency. 


\subsubsection{Unguided Flow Matching}
Flow matching may be \textit{unguided} or \textit{guided}. We use the term `guiding' to refer to conditioning the vector field on additional information, to distinguish it from sample-conditioning, where the vector field is conditioned on a single target sample. In unguided flow matching, the network learns a vector field for a fixed target distribution, while in the guided case, the vector field is conditioned on additional information, such as a text prompt in image generation.

The training objective in \textit{unguided} flow matching is defined as the average mean squared difference between the parametrized vector field $u_\tau^\phi(\bm\theta)$, and the conditional vector field $u_\tau^{\rm target}(\bm\theta_\tau\mid \bm \theta_1)$, which describes the flow from the initial distribution towards individual target samples. An important result in flow matching is that although the loss is defined only in terms of the conditional vector field, optimizing the loss function with respect to this conditional field, which is tractable to compute, also implicitly learns the \textit{marginal} vector field that induces the global flow from $p_{\rm init}$ to $p_{\rm target}$, despite this object being intractable to compute explicitly. 

\begin{figure}
    \centering
    \includegraphics[width=\linewidth]{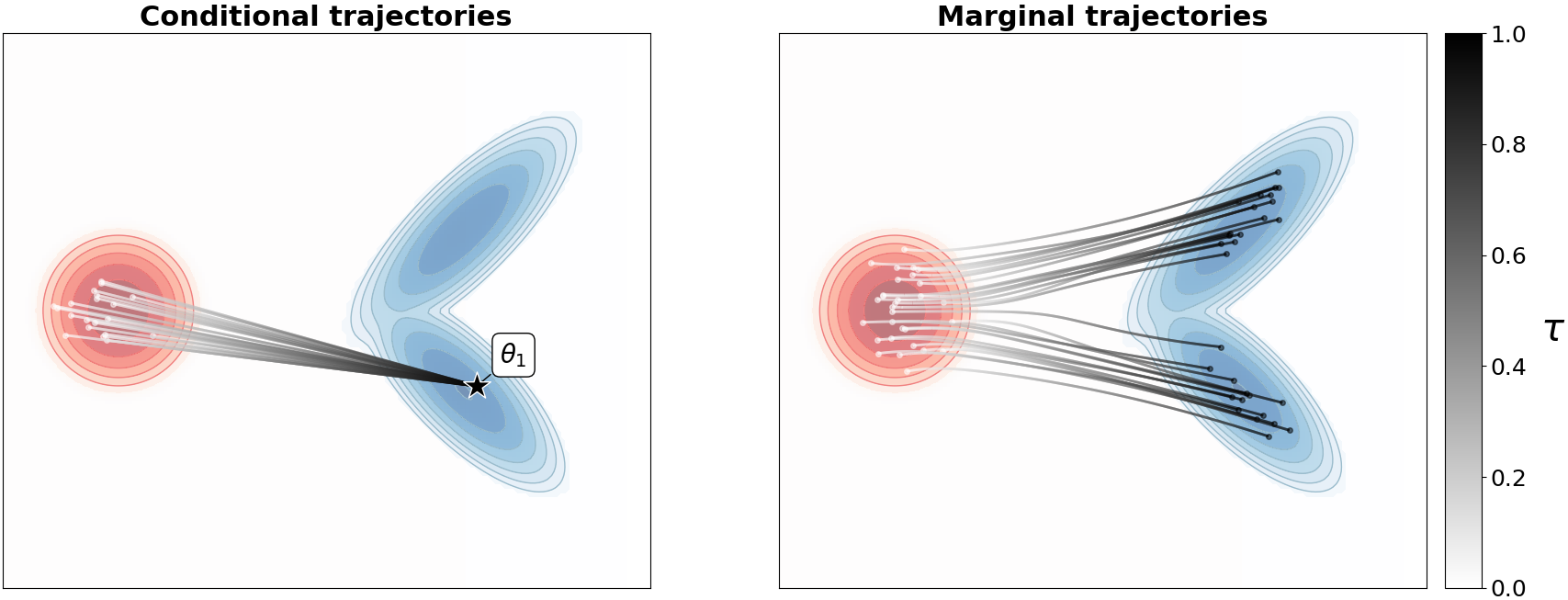}
    \caption{Difference between conditional (left) and marginal (right) trajectories, induced by the conditional and marginal vector field respectively through numerical integration. $p_{\rm init}$ is shown in red, $p_{\rm target}$ is shown in blue.}
    \label{fig:marginal-conditional-trajectories}
\end{figure}

To illustrate their difference, figure \ref{fig:marginal-conditional-trajectories} shows different trajectories generated by a conditional and marginal vector field. Trajectories induced by the conditional vector field map randomly sampled points from the initial distribution to a single target sample specified in advance, whereas trajectories induced by the marginal vector field show a global flow from the initial distribution to the target distribution. Learning the marginal vector field therefore enables sampling from the target distribution itself rather than generating trajectories toward a single predefined point. 

The loss function in unguided flow matching is summarized as
\begin{equation}
    L(\phi)_{\rm unguided}^{FM}=\mathbb{E}_{\tau, \bm\theta_1, \bm\theta_\tau}\|\bm u_\tau^\phi(\bm\theta) - \bm u_\tau^{\rm target}(\bm\theta_\tau\mid \bm \theta_1) \|^2\;,
    \label{eq:unguided-loss}
\end{equation}
where $\mathbb{E}_{\tau, \bm\theta_1, \bm\theta_\tau}$ denotes that the loss is averaged over multiple samples of time step $\tau$, target sample $\bm\theta_1$, and the corresponding interpolation between $\bm\theta_0$ and the target $\bm\theta_1$, $\bm\theta_\tau$. These variables are sampled as follows:
\begin{align}
    \tau &\sim p(\tau) \;\;\;\;0 \leq \tau \leq 1\\
    \bm\theta_1 &\sim p_{\rm target} \\
    \bm\theta_\tau &\sim p_\tau(\bm\theta_\tau\mid\bm\theta_1)
\end{align}
Time can be sampled from an arbitrary distribution defined on $[0,1]$. Often this is $U(0,1)$, but sometimes distributions with more density near $\tau=1$ are preferred \citep{dax2023flowmatchingscalablesimulationbased}. $\bm\theta_\tau$ is sampled from the conditional probability path, $p_\tau$. A conditional probability path is a time-dependent distribution constructed to be equivalent to the initial distribution at $\tau=0$ and converging to a narrow distribution centered on the target $\bm\theta_1$ as $\tau \rightarrow 1$. Sampling interpolated states $\bm \theta_\tau$ from the conditional probability path ensures that these states lie along valid trajectories from the initial distribution to target samples $\bm \theta_1$.

There exist multiple probability paths that satisfy these conditions, such as the family of Gaussian probability paths 
\begin{equation}
    \label{eq:gaussian-path-general}
    p_\tau(\bm\theta \mid \bm\theta_1) = \mathcal{N}(\bm\theta \mid\bm\mu_\tau, \sigma_\tau^2I_n)\;,
\end{equation}
with $I_n$ the identity matrix of size $n$ and where $\bm\mu_\tau$ and $\sigma_\tau$ are chosen such that $p_0=p_{\rm init}=\mathcal{N}(0,I_n)$, and $p_1(\bm\theta\mid \bm\theta_1) = \mathcal{N}(\bm \theta\mid \bm \theta_1, \sigma_{\rm min}^2I_n)$, with $\sigma_{\rm min} \ll 1$, a hyperparameter that ensures the width of the final distribution of the conditional probability path is non-zero for numerical stability. A popular choice for $\bm\mu_\tau$ and $\sigma_\tau$, used in this study as well, is the optimal transport path
\begin{align}
    \label{eq:gaussian-OT-path-mu-sigma}
    \bm\mu_\tau &= \tau \bm\theta_1 \;, \;\;\;\;\;\sigma_\tau = 1-(1-\sigma_{\rm min})\tau\;,
\end{align}
where the mean and standard deviation change linearly with time. 

The corresponding conditional vector field and interpolation are given by 
\begin{align}
    \bm{u}_\tau(\bm{\theta} \mid\bm{\theta}_1) &= \bm{\theta}_1 - (1-\sigma_{\rm min})\bm{\theta}_0 \label{eq:cond-VF-used} \\ 
    \bm{\theta}_\tau &= \bm\mu_\tau + \bm\theta_0\sigma_\tau \notag \\
    &= \tau\bm{\theta_1} + (1-(1-\sigma_{\rm min})\tau)\bm{\theta_0} \label{eq:interpolation-used}
\end{align}
with $\bm{\theta}_0 \sim \mathcal{N}(0,I_n)$ and $\bm{\theta}_1 \sim p_{\rm target}$.

\subsubsection{Guided Flow Matching for FMPE}
Unguided flow matching learns to transform a simple distribution into a complex one, without incorporating additional information related to the target. For posterior estimation however, we would like to include (simulated) observations since the posterior distribution is conditioned on them by definition. We therefore use guided flow matching, where guidance is added to a network by appending additional information $\bm y$ to its input. In contrast to unguided flow matching, each target is now paired with information $\bm y$ such that ($\bm \theta_1, \bm y$) are drawn from a joint distribution. The guided flow matching loss is then given by 
\begin{equation}
\begin{split}
    L(\phi)_{\rm guided}^{FM}=&\mathbb{E}_{\tau\sim p(\tau),\; \bm\theta_1,\bm y\sim p(\bm \theta_1, \bm y),\; \bm \theta_\tau\sim p_\tau(\bm \theta \mid \bm \theta_1)}\\ &\left[\|\bm u_\tau^\phi(\bm\theta_\tau\mid\bm y) - \bm u_\tau^{\rm target}(\bm \theta_\tau\mid\bm \theta_1) \|^2\right]\;,
\end{split}
\end{equation}
which differs from the unguided loss in equation \ref{eq:unguided-loss} only through the addition of $\bm y$ to the network input and the joint sampling of target $\bm \theta_1$ and label $\bm y$. 

To adapt guided flow matching for posterior estimation, the following has been proposed in \citet{dax2023flowmatchingscalablesimulationbased}: The labelled targets that are usually sampled from a joint distribution $\theta,y \sim p(\theta,y)$ can instead be generated using $\bm\theta_1, \bm y \sim p(\bm\theta) p(\bm y\mid\bm\theta)$ by applying Bayes' theorem. This means parameters are sampled from the prior, and a simulation is subsequently generated from them. The adapted training target becomes
\begin{equation}
\begin{split}
   L(\phi)^{FMPE}_{\rm guided} = &\mathbb{E}_{\tau\sim p(\tau),\; \bm\theta_1, \bm y\sim p(\bm\theta)p(y\mid \bm\theta_1), \;\bm\theta_\tau\sim p_\tau(\bm\theta \mid\bm\theta_1)} \\ &\left[ \|u_\tau^\phi (\bm\theta_\tau\mid\bm y) - u_\tau^\text{target} (\bm\theta_\tau\mid\bm\theta_1)\|^2 \right]  \;.
\end{split} 
\label{eq:FMPE-loss}
\end{equation}

\subsection{Trans-Dimensional Flow Matching Posterior Estimation}
In order to solve the trans-dimensional inference problem at hand, we extend FMPE to posterior distributions that span multiple dimensions. Instead of generating samples of fixed size, with t-FMPE we can generate trans-dimensional samples for any observed light curve $\{y_k\}$,\footnote{Assuming its true parameters are supported by the prior.} without requiring retraining and without needing to specify $N$ in advance. This section describes the method in detail: how the vector field is parametrized, how the training objective in equation \ref{eq:FMPE-loss} is modified and how training steps and inference are performed. 

\subsubsection{Network architecture}
 The network is designed to represent a vector field $\bm{u}_{\tau}^\phi$ whose dimensionality varies with $N$. As mentioned in section \ref{sec: training-data}, a maximum number of components $N_{\rm max}$ is set in advance, such that ($\bm{u}_{\tau}^\phi$, $\bm \theta$) $\in {\rm I\!R}^{4\times N_{\rm max}}$. Any excess parameter components $\bm \theta_{n>N_{\rm true}}$ are masked out during training, based on the true number of components for each simulated light curve (teacher forcing). After training, the number of components is assumed unknown and must be inferred by the network. Therefore, the network is divided into two parts: a classifier, which predicts the number of components $N$, and a transformer encoder that represents the corresponding vector field $\bm{u}^\phi_\tau$. These modules do not share any parameters, but are trained in parallel with a single optimizer instance. 
 
Given the complexity of the network setup, the following sections provide a general conceptual overview of the method, while specific implementation details such as training settings are presented separately in section \ref{sec:network-configurations} and appendix \ref{sec:implementation-details}.
 
\begin{figure}
    \centering
    \includegraphics[width=\linewidth]{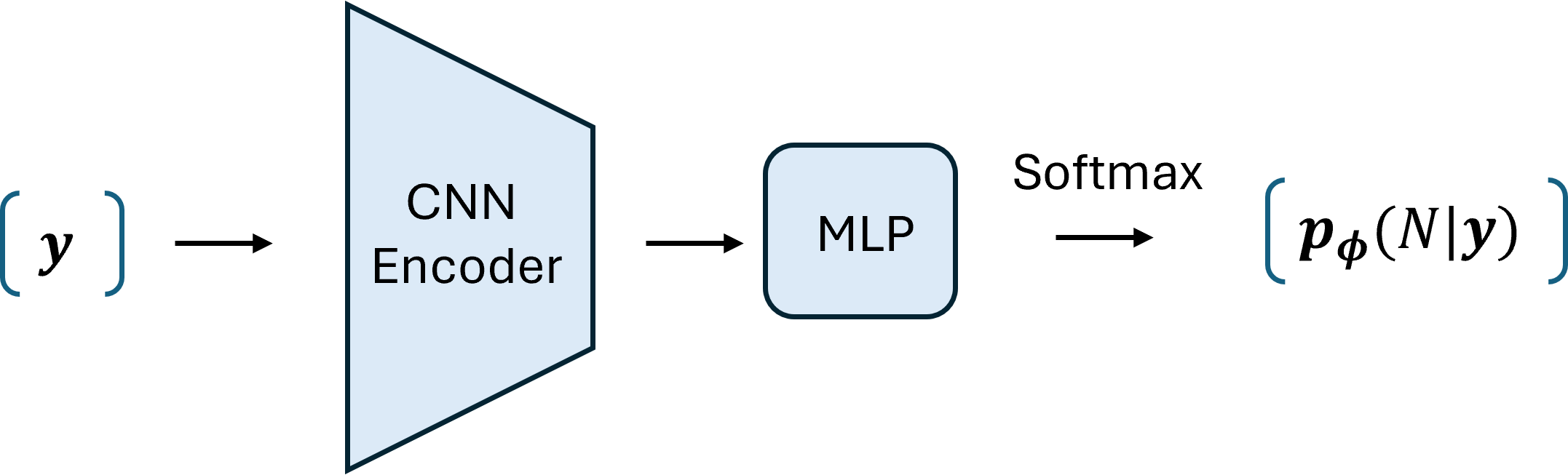}
    \caption{Diagram of classifier architecture. Note that the classifier uses a separate CNN encoder that is independent from the one used in token generation (figure \ref{fig:token-creation-diagram}). }
    \label{fig:classifier-diagram}
\end{figure}

 \paragraph{Component Classifier}
The component classifier estimates the posterior over the number of components. Given a light curve $\bm y \in {\rm I\!R}^K$, it gives a probability vector $\bm{p}_\phi(N\mid y)$ of length $N_{\rm max}$, where each entry corresponds to the probability that $\bm y$ contains $N \in [1\dots N_{\rm max}]$ components. As shown in figure \ref{fig:classifier-diagram}, it consists of a one-dimensional multi-scale CNN embedding network followed by a multi-layer perceptron (MLP) \citep{OG-multi-scale-cnn, multi-scale-cnn-review}. The embedding network compresses the input $\bm y\in {\rm I\!R}^K$ into an encoded representation $\bm y_e \in {\rm I\!R}^{d_y}$ with $d_y$ the predefined size of $\bm y_e$. This encoding is needed as an MLP on its own has no translation invariance and would treat each timestep as independent, whereas the embedding network applies convolutions and pooling to extract temporal features from the time series in a stable and translation-invariant way \citep{multi-scale-cnn-review}. A Softmax activation is applied to the output layer to ensure that 
 \begin{equation}
     \sum_{j=1}^{N_{\rm max}} p_{j,\phi}(N\mid y) =  1\;.
 \end{equation}

Further implementation details, such as the exact MLP an CNN layer configurations, are provided in appendix \ref{sec:implementation-details}. 

\paragraph{Transformer Encoder}\label{sec:transformer-encoder-method-small}
To parametrize the vector field $\bm{u}_\tau^\phi(\bm{\theta}\mid \bm{y}, N)$, a transformer encoder is used. Transformers are able to take into account long range context through self-attention, scale well, and can accommodate the trans-dimensional aspect through masking of excess tokens. This makes them preferable to convolutional or recurrent neural networks, which often struggle to capture long-range correlations and provide no obvious way for handling inputs of variable effective size. For a more in-depth overview of transformers and multi-headed self-attention, we refer to \citet{vaswani2023attentionneed, allam2023_paying_attention_astronomical_transients, hands-on-llms-book, tanoglidis-2023-transformersscientificdatapedagogical}.

\begin{figure}
    \centering
    \includegraphics[width=\linewidth]{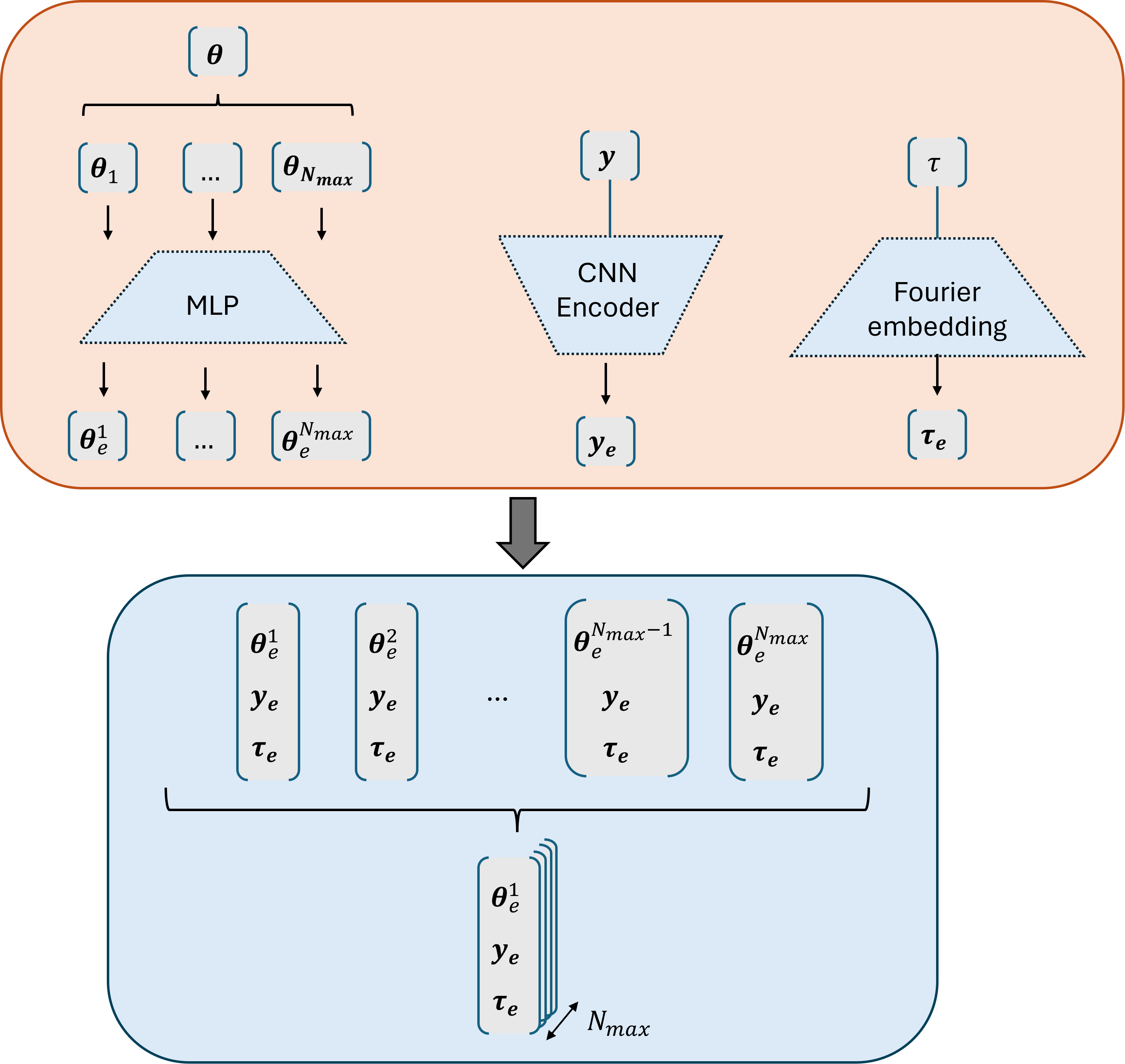}
    \caption{Diagram showing how input tokens are created. Parameters $\theta$, time series $y$ and time $\tau$ are encoded and concatenated into $N_{\rm max}$ tokens. Each token represents a burst component that may or may not be included in self-attention.}
    \label{fig:token-creation-diagram}
\end{figure}

\paragraph{Embedding and Tokenization}
Transformers require tokenized inputs, which means any input must be divided into segments, each translated to some vector embedding of fixed dimension $d_t$ to create a token $T_i \in {\rm I\!R}^{d_t}$. For large language models, this is done by dividing sentences into words or parts of words and transforming these to a vector embedding. In our case, tokens are created by embedding and concatenating three variables: the parameter vector of the $n$-th component at time $\tau$ $\bm{\theta}_{n, \tau}$, flow matching time $\tau$ and the time series $\bm{y}$. As illustrated in figure \ref{fig:token-creation-diagram}, this results in $N_{\rm max}$ tokens, such that each token represents one burst component. 

Since the raw inputs vary considerably in dimension: $\bm{\theta}_{n, \tau} \in {\rm I\!R}^{4}$, $\tau\in {\rm I\!R}^1$ and $\bm{y}\in{\rm I\!R}^{1000}$, they are either compressed or expanded through embedding networks as shown in figure \ref{fig:token-creation-diagram}. The resulting embeddings are then concatenated to form $N_{\rm max}$ tokens. Each token has a size of $d_t = d_y + d_{\theta_{\tau}} + d_\tau$, with $d_y : d_{\theta_\tau} : d_\tau = 2: 1:1$, meaning the sizes of the embedded vectors $\bm y_e, \bm \theta_e$ and $\bm \tau_e$ always follow this ratio respectively. This is done to attribute more weight to the data $\bm y$. The exact values depend on the network settings and are given in table \ref{tab:network-configurations}.

The decomposed parameter vectors $\{\bm\theta_n\}_{n=1}^{N_{\rm max}}$ are encoded by a Multi-Layer Perceptron with layer dimensions [4, 16, 32, 32, 64, 64, $d_\theta$], with SiLU activation \citep{misra_2020_mish_silu_activation}, defined as SiLU$(x)=x\cdot \sigma(x)$, except on the output layer. 

The flow matching time $\tau$ is embedded with a fixed Fourier embedding, 
\begin{align}
    \omega_j &= 10^\frac{3j}{d_\tau/2 - 1} \;\; \text{with }\;j=0, 1\dots d_\tau/2 -1\\
    \bm{\tau}_e &= \left\{\sin(\omega_j\tau),\;\cos(\omega_j\tau)\right\}_{j=0}^{d_\tau / 2}.
\end{align}
This type of embedding is commonly used to map low-dimensional or scalar input to higher-dimensional input in a way that is able to capture high-frequency signals from data \citep{tancik2020fourierfeaturesletnetworks}. In this setting, this means even small differences in $\tau$ result in substantial changes in $\bm{\tau}_e$, such that the final model will be sensitive to small-scale variations in $\tau$. 

The time series $\bm y$ is embedded with an embedding network, identical in architecture to the one used by the classifier. These two networks are optimized entirely independently, as each is associated with a distinct objective (described in section \ref{sec:loss-and-training}).



\begin{figure}
    \centering
    \includegraphics[width=0.9\linewidth]{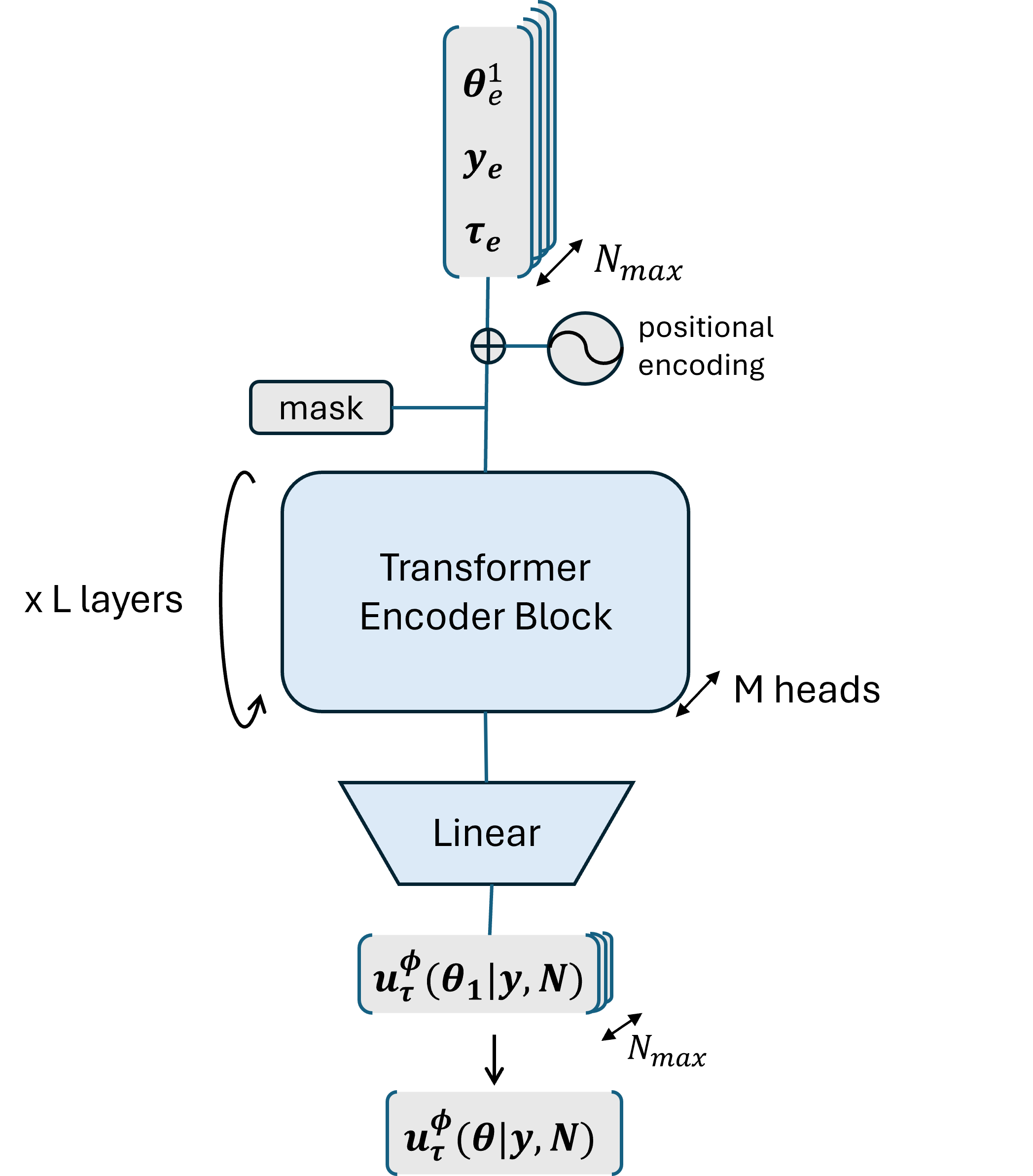}
    \caption{Diagram with a high-level view of the transformers architecture, further explained in section \ref{sec:transformer-fw-pass}. Contributing tokens are selected with a mask that depends on the true number of components (during training).}
    \label{fig:transformer-diagram}
\end{figure}

\begin{figure}
        \centering
        \includegraphics[width=0.5\linewidth]{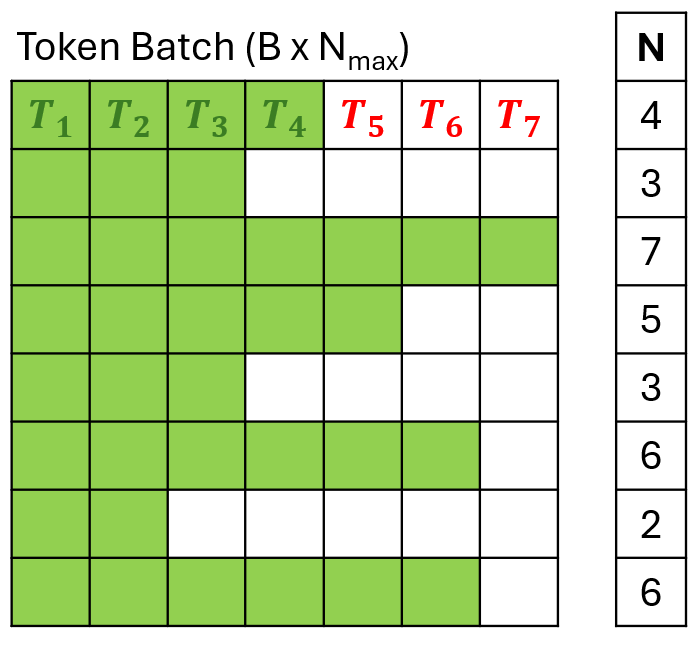}
        \caption{Diagram of the mask that is used to select relevant tokens for an example batch of $B=8$ samples with $N_{\rm max}=7$. Column on the right indicates the number of components provided, rows on the left indicate the corresponding selected token positions in green. }
        \label{fig:token-mask}
\end{figure}

\paragraph{Masking and Forward Pass}\label{sec:transformer-fw-pass}
The forward pass through the transformer encoder that follows embedding and tokenization is shown in figure \ref{fig:transformer-diagram}. During self-attention, only tokens $T_{n\leq N}$ participate. This is done by applying a mask, illustrated in figure \ref{fig:token-mask}, that masks out non-contributing tokens corresponding to the number of components $N$. During training, this mask is created with the true number of components in the light curve $N\gets N_{\rm true}$.\footnote{This is known as 'teacher forcing` in deep learning}. At inference, $N$ is sampled from the classifier instead $N\sim \bm{p}_{\phi}(N\mid \bm{y})$ 

After applying the mask, tokens are passed through $L$ encoder blocks, each with $M$ self-attention heads. The final output of the encoder blocks has shape $B \times d_t \times N_{\rm max}$, where $B$ denotes the batch size and $d_t$ the token dimension. A final linear layer maps this to $B \times 4 \times N_{\rm max}$, corresponding to the conditional vector fields of each burst component. These outputs are then concatenated into a single vector of dimension $B \times 4N_{\rm max}$, representing the vector field $\bm{u}_\tau^\phi$.  

\subsubsection{Loss and Training Procedure}\label{sec:loss-and-training}
During training, we aim to approximate the marginal vector field that, guided by a light curve $\bm{y}$, induces a flow from the base distribution to the posterior. This is done by optimizing the network with respect to the conditional vector field $\bm{u}_\tau(\bm{\theta}\mid \bm{\theta}_1)$ defined in equation \ref{eq:cond-VF-used}. Note that although the conditional vector field itself is independent of time $\tau$, the learned marginal vector field will be time-dependent. 

The network parameters $\phi$ are optimized with respect to the following batch-averaged loss function, 
\begin{equation}
\begin{split}
    L(\phi) = E_{\tau, \bm{\theta}_1, \bm{y}, \bm{\theta}_\tau, N_{\rm true}}&\Big[ \big\| \bm{u}_\tau^\phi(\bm\theta_\tau\mid \bm{y}, N_{\rm true}) - (\bm\theta_1 - (1-\sigma_{\rm min})\bm\theta_0) \big\|^2 \\ &- \log{p_\phi(N_{\rm true}\mid \bm{y})}\Big]  
\end{split}
\label{eq:loss-voluit}
\end{equation}
over parameters that are sampled as follows: 
\begin{align}
    \tau &\sim p(\tau), \;\;p(\tau)\propto \tau^\alpha \\
    \bm\theta_1, N_{\rm true}, \bm y &\sim p(\bm\theta)p(N)p(\bm y\mid \bm \theta, N_{\rm true}) \\
    \bm \theta_\tau &\sim p_\tau(\bm \theta\mid \bm\theta_1)\;, 
\end{align}
with $p_\tau(\bm \theta \mid \bm\theta_1)$ the linear Gaussian conditional probability path (equation \ref{eq:gaussian-path-general}). $\tau$ is sampled from a power law distribution with $\alpha=\frac{4}{3}$, inspired by \citet{dax2023flowmatchingscalablesimulationbased}, who noted that the vector field tends to get more complex as $\tau \rightarrow 1$, making it useful to focus more on that regime during training. This is done through inverse sampling. 

The loss contains two contributions. The first is the mean square error (MSE) between the predicted and conditional vector field, the second is the multi-class cross-entropy loss (CEL) from the classifier
\begin{align}
    CEL &= -\sum_{n=1}^{N_{\rm max}} \delta_{n, N_{\rm true}}\log p_\phi(n\mid y) \notag \\
    &= - \log{p_\phi(N_{\rm true}\mid y)}\;,
\end{align}
where 
\begin{equation}
    \delta_{n, N_{\rm true}} = \begin{cases}
        1 \;\;\;\text{ if } n = N_{\rm true}  \\
        0 \;\;\;\text{ otherwise}
    \end{cases}\;.
\end{equation}
Since the MSE and CEL contributions were found to be of the same order of magnitude during training, no loss weighting was applied. Contributions to the MSE loss from irrelevant components $\{\bm u^\phi_{\tau}\}_{n>N_{\rm true}}$ are masked out. One training step is summarized in algorithm \ref{alg:training-step}. 

{\centering
\begin{minipage}{0.9\linewidth}
\renewcommand{\footnoterule}{}
  \begin{algorithm}[H]
    \begin{algorithmic}[1]
    \caption{Training step}\label{alg:training-step}
    \State $\bm\theta_1\sim p(\bm\theta)$ 
    \State $N_{\rm true}\sim p(N)$
    \State Generate batch of simulated bursts $\bm y \sim p(\bm y\;|\;\bm\theta_1, N_{\rm true})$
    \State $\bm\theta_0\sim\mathcal{N}(0,I_n)$
    \State $\tau \sim p(\tau)$ 
    \State Find interpolation $\bm\theta_\tau=$ eqn \ref{eq:interpolation-used}
    \State target = $\bm{u}_\tau(\bm\theta \mid \bm\theta_1) =$ eqn \ref{eq:cond-VF-used}
    \State prediction = $\bm{u}_\tau^\phi =$ Network$(\bm\theta_\tau, \tau, \bm{y}, N_{\rm true})$
    \State loss = MSE$($prediction, target$)$ + CEL 
    \State Update network parameters with SGD (\texttt{Adam}; \citep{kingma2017_adam_methodstochasticoptimization})
    
\end{algorithmic}
\end{algorithm}
\end{minipage}
\par
}

\subsubsection{Inference}\label{sec:FM-inference}
At inference, posterior samples of varying dimension are drawn for a given observational or synthetic light curve $\bm y$. First, the number of components is sampled from the classifier $N \sim \bm p_\phi(N\mid \bm y)$ and the initial state is sampled from the base distribution $\bm\theta_0 \sim \mathcal{N}(0,I_n)$. Then, the ODE defined by $\bm u_\tau^\phi(\bm\theta\mid \bm y, N)$ is solved numerically with an Euler integration scheme (equation \ref{eq:Euler_FM}). The initial condition is set $\bm\theta=\bm\theta_0$ at $\tau=0$, and integration proceeds for $n_{\rm steps}=200$ steps to produce a posterior sample. The described procedure is summarized in algorithm \ref{alg:inference}. To generate an approximate posterior distribution, many samples are integrated in parallel. 

{\centering
\begin{minipage}{.9\linewidth}
  \begin{algorithm}[H]
    \begin{algorithmic}[1]
    \caption{Generating an FM posterior sample}\label{alg:inference}
    
\State $N \sim \bm p_\phi(N\mid \bm y)$. 
\State $\bm \theta_0 \sim \mathcal{N}(0, I_n)$
\State $\Delta\tau = 1 / n_{\rm steps}$
\Statex \textbf{Integrate:}
\While{ $\tau \neq 1$}
    \State $\bm\theta_{\tau+1} \gets \bm\theta_\tau + \bm u^\phi_\tau(\bm\theta_\tau \mid \bm y, N)\Delta\tau$ 
       \State $\tau \gets \tau + \Delta\tau$
\EndWhile

\Statex \textbf{Result: }$\bm\theta_1 \sim p^\phi(\bm \theta\mid \bm y)$, $N \sim \bm p_\phi(N\mid y)$

\end{algorithmic}
  \end{algorithm}
\end{minipage}
\par
}

\subsubsection{Network Configurations and Training Settings}\label{sec:network-configurations}
In this study, a total of four networks are trained for t-FMPE, differing in either architecture or training data settings. Initially, networks were trained with $N_{\rm max}=5$ and $N_{\rm max}=10$ to verify the performance on simulated data, while allowing meaningful exploration of the already high-dimensional posterior. For inference on observational X-ray and radio bursts, which included more complex light curves, networks were trained with $N_{\rm max}=20$ for twice the amount of training steps to accommodate the increased complexity. These two networks differ only in the noise distribution of the training data, since X-ray and radio data enjoy different noise distributions (see section \ref{sec:adding noise}). A summary of the network configurations is given in table \ref{tab:network-configurations}. 

\begin{table}
\caption{Overview of the different network configurations and training settings used. Network size is given in number of parameters.}
    \centering
    \begin{tabular}{|m{5.7em}|c|c|c|c|}
\toprule
\bfseries Network & 1 & 2 & 3 & 4 \\
\midrule
Inference on & Simulations & Simulations & X-ray data & FRB data \\
Signal noise & Poisson & Poisson & Poisson & Gaussian \\
$\lambda_{\rm bkg}$ & 5 & 5 & 3 & 0 \\
$N_{\rm max}$ & 5 & 10 & 20 & 20 \\
$L$ & 6 & 12 & 12 & 12 \\
$M$ & 4 & 8 & 8 & 8 \\
$d_y$ & 64 & 128 & 128 & 128 \\
Training steps & 300k & 300k & 600k & 600k \\
Parameters & 4.6M & 12.2M & 12.2M & 12.2M \\
Training time & 5h 19m & 10h 16m & 28h 51m & 28h 47m \\
Inference time 1000 samples & 2.0s& 5.0s& 10.0s & 10.0s\\
\bottomrule
\end{tabular}
    \label{tab:network-configurations}
\end{table}
\label{sec: training-imp-details}
All networks are trained with a batch size of 1024 on an NVIDIA A100 GPU with an \texttt{Adam} optimizer. For stability, gradient norms of the network parameters are clipped at 1 using \texttt{torch.nn.utils.clip\_grad\_norm\_}.




\section{Results}
This section presents inference results from applying the trained t-FMPE networks to simulated light curves, as well as observed magnetar and fast radio bursts. Using simulated data, we compare results against reference posteriors obtained through MCMC sampling and evaluate the classifier. This is followed by trans-dimensional inference results of an X-ray burst from magnetar SGR J1550-5418 which we compare to trans-dimensional nested sampling results obtained using the method developed by \citep{Huppenkothen_2015}. Finally, we apply the t-FMPE method to three multi-peaked FRBs. 

\subsection{Results with Simulated Data}\label{sec: results-sim-data}
The results in this section stem from FM networks 1 and 2 as detailed in table \ref{tab:network-configurations}. Network 1 is used in the following two sub-sections for fixed-$N$ comparisons against MCMC results for a lightcurve with $N_{\rm true} =1$ and $N_{\rm  true}= 5$. Therefore, results from the classifier are not incorporated in the sampling process at this stage, as $N$ is assumed to be known. This is done by only selecting FM posterior samples where $N=N_{\rm true}$. Network 2, trained on light curves with up to $N_{\rm max}=10$ components, is used to validate the classifier.

\subsubsection{Comparison for Singly-Peaked Burst}

\begin{figure}
    \centering
    \includegraphics[width=0.8\linewidth]{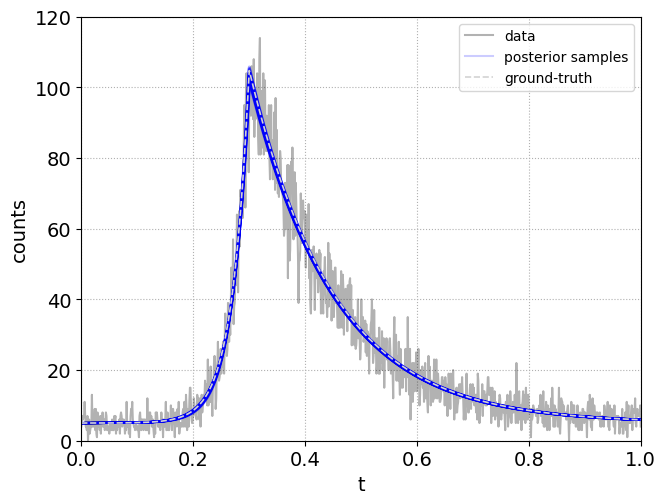}
    \caption{FM posterior predicted light curves (blue) for a simulated flux light curve (grey). The dotted line indicates the ground-truth noise-less flux, with true values $t_0=0.3$, $r=3\cdot 10^{-2}$, $s=5$, $A=100$. }
    \label{fig:sampled-curves-1-peak-mcmc-fm}
\end{figure}

\begin{figure}
    \centering
    \includegraphics[width=0.9\linewidth]{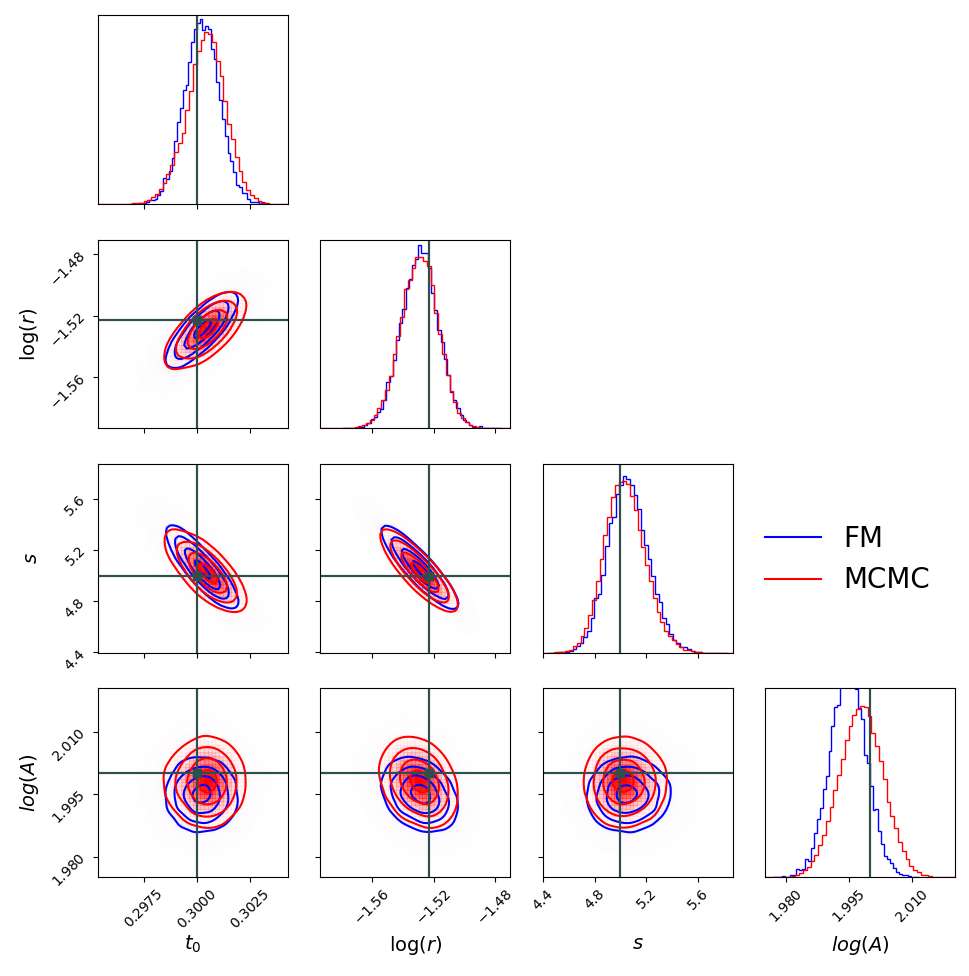}
    \caption{Posterior distribution comparison between MCMC (red) and FM (blue) for the simulated flux sample with $N=1$ peak shown in figure \ref{fig:sampled-curves-1-peak-mcmc-fm}. The FM model was trained on data with $N_{\rm max}=5$. For MCMC, the true number of components was given. The contours shown are the [0.5, 1, 1.5, 2]-$\sigma$ contours. $\lambda_{bkg}=5$. The true parameter values are $t_0=0.3$, $\log(r)=-1.52$, $s=5$, $\log(A)=2$.}
    \label{fig:corner-fixed-1-peak-fm-vs-mcmc}
\end{figure}

A simulated burst with a single peak was generated, shown in figure \ref{fig:sampled-curves-1-peak-mcmc-fm}, with Poisson noise, $\lambda_{\rm bkg}=5$ and true parameter values $t_0=0.3$, $r=3\cdot 10^{-2}$, $s=5$, $A=100$. Posterior samples were produced for this burst using the MCMC and FM procedures described in sections \ref{sec:mcmc-method} and \ref{sec:FM-inference}. The MCMC sampler used 40 walkers (chains) which converged after 4500 steps, generating $1.8 \cdot 10^5$ posterior samples in 30 seconds. FM inference took 40 seconds for $2.5 \cdot 10^4$ samples. Because the estimated autocorrelation time of the MCMC samples is relatively long ($\langle\tau\rangle=43.5$, with the longest estimate being $\tau_{\rm max}=44.4$), a larger number of sampling steps is required in MCMC to obtain a sufficient amount of independent posterior samples. 

Figure \ref{fig:corner-fixed-1-peak-fm-vs-mcmc} compares the resulting FM and MCMC posterior distributions constructed from these samples in a corner plot. This corner plot visualizes the posterior using the one- and two-dimensional marginal distributions of the four peak parameters: $t_0$, $r$, $s$ and $A$. The MCMC and FM distributions show substantial overlap, both including the true parameter values. To quantify their similarity, a Classifier Two-Sample test (C2ST) was performed \citep{lopez-paz-2018-revisiting-c2st}. A C2ST quantifies the difference between two distributions by training a binary classifier to distinguish between their samples. Indistinguishable distributions produce around 50\% classification accuracy, while substantially different distributions approach 100\%. For the distributions in figure \ref{fig:corner-fixed-1-peak-fm-vs-mcmc}, the resulting C2ST score was 54.8\%.

The peak time $t_0$, skewness $s$ and rise time $r$ exhibit clear linear relationships in the corner plot. This follows from skew and rise time being inversely related through the fall time, $f = s r$. The linear relation between parameter pairs  ($t_0$, $s$) and ($t_0$, $r$) logically follows from compensating for a shifted peak location. For example, when the estimated peak time is off-set to the right, the corresponding fall time should decrease to stay consistent with the observed flux, lowering the skewness.  

To further validate our results, we check whether samples drawn from the learned posterior generate models that are consistent with observations. To this end, noise-free light curves are generated from posterior samples $\bm \theta$ and overlaid on noisy data. Figure \ref{fig:sampled-curves-1-peak-mcmc-fm} shows 100 FM posterior predicted light curves overlaid on the simulated and ground-truth flux, demonstrating good agreement. 

\begin{figure}
    \centering
    \includegraphics[width=\linewidth]{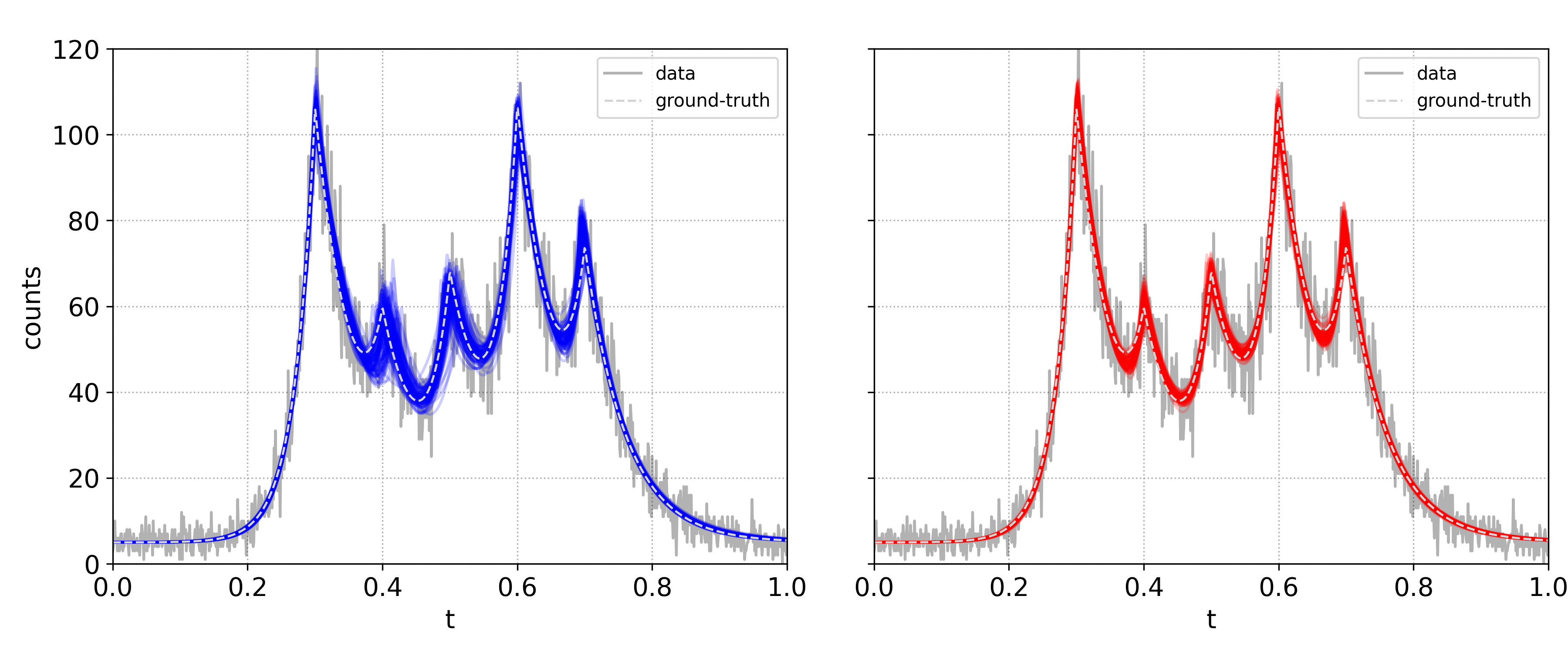}
    \caption{100 FM (left) and MCMC (right) noise-free posterior predicted light curves for the simulated flux shown in gray with $N=5$ components. The dashed line indicates the ground-truth, noise-less flux. }
    \label{fig:samples-5-peaks-mcmc-fm}
\end{figure}

\subsubsection{Comparison for Five Peaks}
We increase the complexity of the inference task by now generating a simulated burst with $N_{\rm true}=5$  well-separated components. As in the previous section, posterior samples were produced through t-FMPE and MCMC. The MCMC sampler took 3.15 hrs to converge, generating $5.328 \cdot 10^6$  posterior samples in the process. FM inference again required 40 seconds for 25k samples using the same trained network as before. Extrapolating, FM inference would theoretically take 2.37 hours to generate an equivalent number of samples. However, this amount is more of a result of the relatively long autocorrelation time in MCMC, with an estimated average of $\langle \tau \rangle = 671$ and maximum of $\tau_{\rm max}=887$, requiring many sampling steps to ensure enough independent samples. This is not an issue in t-FMPE, where reliable approximate posteriors can generally be constructed from far fewer samples. Figure \ref{fig:samples-5-peaks-mcmc-fm} compares 100 resulting FM and MCMC noise-free posterior predicted light curves. All light curves appear consistent with the simulated flux, though the FM predictions show a wider spread. A corner plot comparison is provided in the Appendix, (figure \ref{fig:5-peaks-corner-mcmc-fm}). The resulting C2ST score of the two distributions was 92\%. This suggests the distributions differ substantially, though overall their shapes and covariances appear to agree. This is likely because C2ST is sensitive to \textit{any} difference in distribution \citep{lopez-paz-2018-revisiting-c2st}. Since the FM posteriors are substantially wider, samples that fall outside of the MCMC posterior range in \textit{any} of the marginals are trivially classified by the MLP as FM samples, leading to the high classification accuracy of 92\%. 

\subsubsection{Classifier Evaluation with Simulated Data}
The average outcome of the classifier from Network 2 is evaluated by using simulated light curves where the number of components is known.  

\begin{figure}
    \centering
    \includegraphics[width=0.95\linewidth]{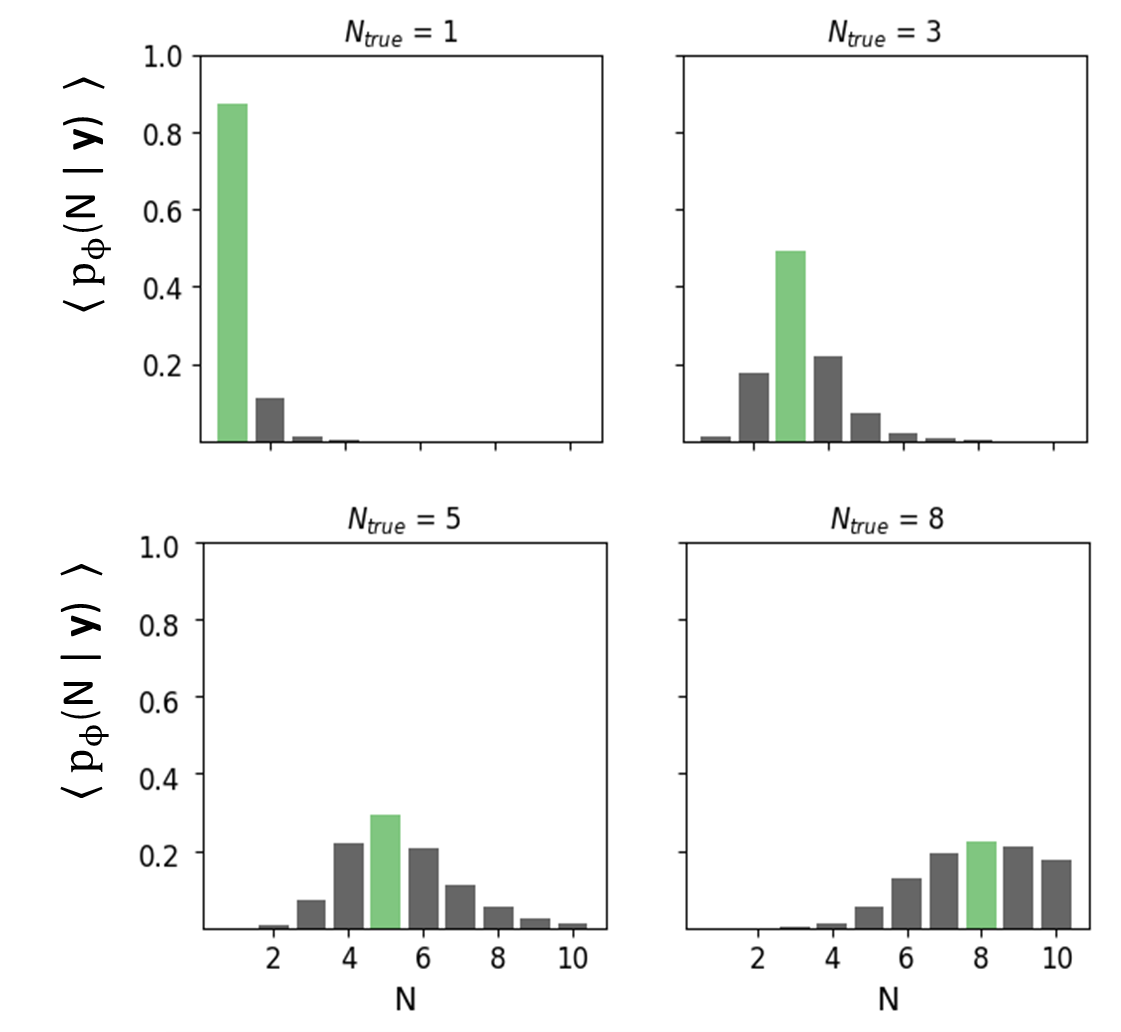}
    \caption{ $p_{\phi}(N\mid \bm y)$ averaged over 2000 simulated light curves $y \sim p(\bm\theta)p(\bm y\mid\bm \theta, N_{\rm true})$ with $N_{\rm true}\in[1,3,5,8]$. $p_\phi(N_{\rm true}\mid \bm y)$ is shown in green.}
    \label{fig:p_N_2000_curves_average}
\end{figure}

For each $N_{\rm true}\in[1,3,5,8]$, 2000 light curves were simulated and their average predicted probability vectors $\{\langle \bm p_\phi(N\mid \bm y)\rangle\}^{N_{\rm max}}_{N=1}$ are illustrated in figure \ref{fig:p_N_2000_curves_average},
where 
\begin{equation}
    \langle \bm p_\phi(N\mid \bm y)\rangle = \frac{1}{S}\sum^{S}_{j=1} \bm p_\phi(N\mid \bm y_j)
\end{equation}
with $S=2000$, $p_\phi$ the predicted component posterior for $\bm y_j$, the $j$-th simulated light curve with $N$ components. The true parameter values of the simulated light curves were drawn from the prior distributions mentioned in table \ref{tab:priors}.

For individual light curves, it is expected for the maximum value of $p(N\mid  \bm y)$ to not always be at $N=N_{\rm true}$, depending on the signal-to-noise ratio (SNR) and component overlap within a given light curve. The number of components is straightforward to predict for light curves with high SNR and peak separability. For individual cases where peaks are obfuscated by noise, where noise may resemble an exponential peak by chance, or where components have a significant degree of overlap, the most probable estimate need not necessarily be $N_{\rm true}$. However, we \textit{do} expect this to be true when averaging over an ensemble of light curves, as is confirmed in figure \ref{fig:p_N_2000_curves_average}, indicating the classifier is predicting components accordingly. 

When more components are contributing to a light curve of the same duration, there is a larger probability of overlap between components, which in turn makes it less trivial for the classifier to deconstruct it into $N_{\rm true}$ components. This is reflected in the increased spread of the probability mass over multiple bins, with less difference between the most probable estimate and the neighbouring values as $N_{\rm true}$ increases. Several causes may be contributing to the ambiguity of light curves, most prominently the amount of overlap between components, i.e. components where the distance between consecutive peaks is small. When $N_{\rm true}$ is large, the probability of overlapping components increases, leading to a larger amount of ambiguous light curves and hence a broader component posterior $\bm p_\phi(N\mid \bm y)$.

\subsection{Inference on Observational Data}\label{sec:OD-results}
Here, the results are presented of inference on observational radio and X-ray bursts observed by CHIME and the Fermi/GBM respectively, as described in section \ref{sec:OD}. From this point on, posterior samples may differ in the corresponding number of components $N$ sampled from the classifier.

\subsubsection{Magnetron Comparison for Magnetar Burst}\label{sec:NS-comparison-results}

We compare FM inference results to those of \texttt{magnetron}, a trans-dimensional nested sampling (t-NS) algorithm that uses the same burst component model \citep{Huppenkothen_2015}. Nested sampling considers a set of `live points' that is first drawn from the prior. At each iteration, the live point with the lowest likelihood is sampled and replaced by a new point, constrained to have a higher likelihood than the one removed. This process shrinks the prior volume and steps through a set of nested likelihood contours as a run progresses. \citet{Huppenkothen_2015} use a variant known as diffusive nested sampling, which softens the hard likelihood constraint of standard nested sampling \citep{diffusive-nested-sampling-2010, nested-sampling-skilling-2004}. They also use trans-dimensional jumps to propose moves that add or remove components, such that the resulting posterior samples may vary in $N$. A more detailed description of this procedure can be found in Section 3 of \citep{Huppenkothen_2015}. 

Here, we compare results for an X-ray burst (burst trigger ID 090122173) observed from SGR J1550-5418 and included in \citet{Huppenkothen_2015}. This burst was specifically chosen to allow for a first comparison of the predicted light curves \textit{and} number of components between two distinct trans-dimensional sampling methods; t-FMPE and t-NS. The t-NS algorithm from \cite{Huppenkothen_2015} was re-run for this burst with the original \texttt{magnetron} code, but with adjusted priors that match those used in t-FMPE.

\begin{figure}
    \centering
    \begin{subfigure}{0.25\textwidth}
        \centering
        \includegraphics[height=1.5in]{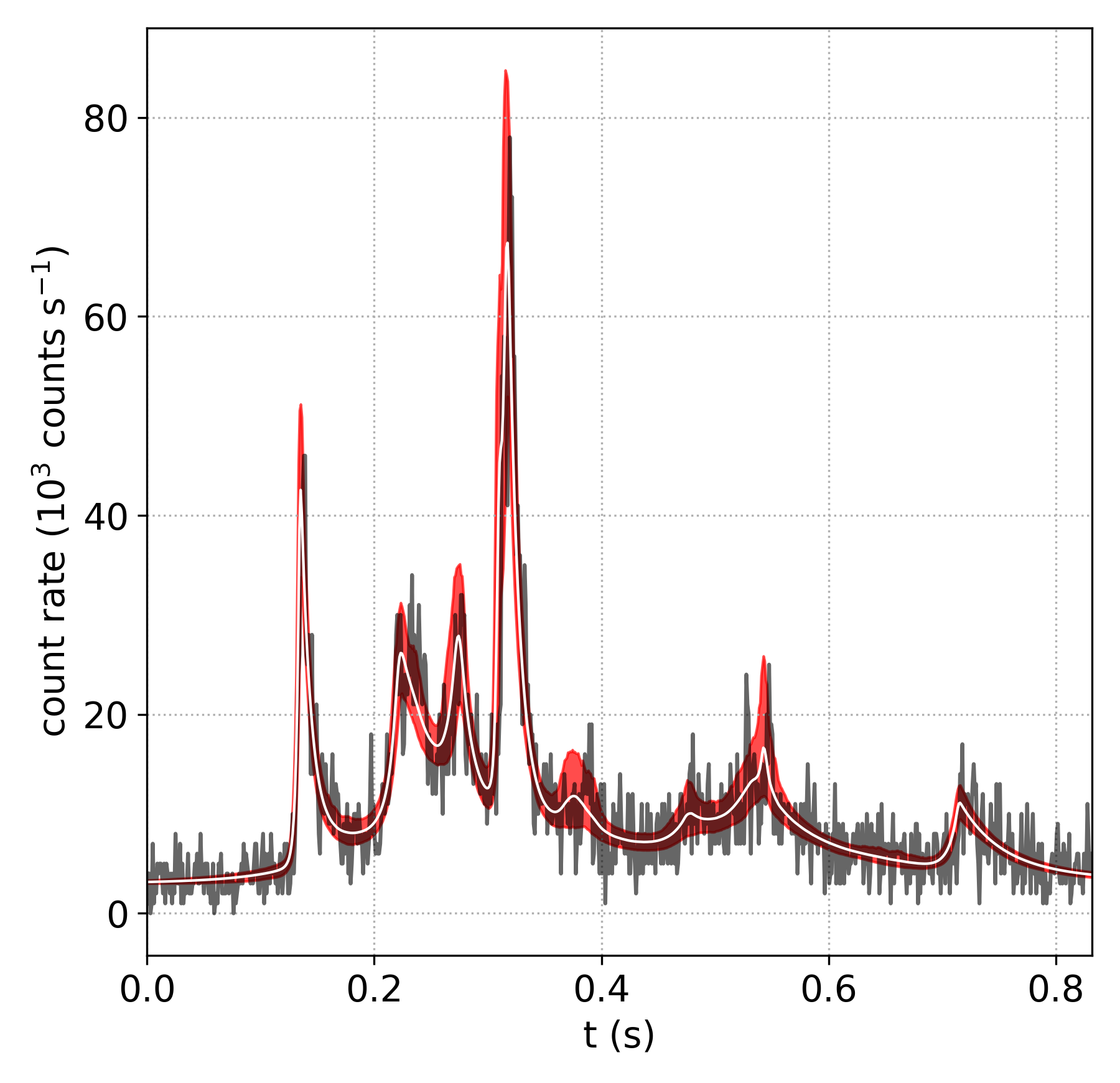}
        \caption{Flow Matching}
        \label{fig:FM-magnetar-paper-plot}
    \end{subfigure}%
    ~
    \begin{subfigure}{0.25\textwidth}
        \centering
        \includegraphics[height=1.5in]{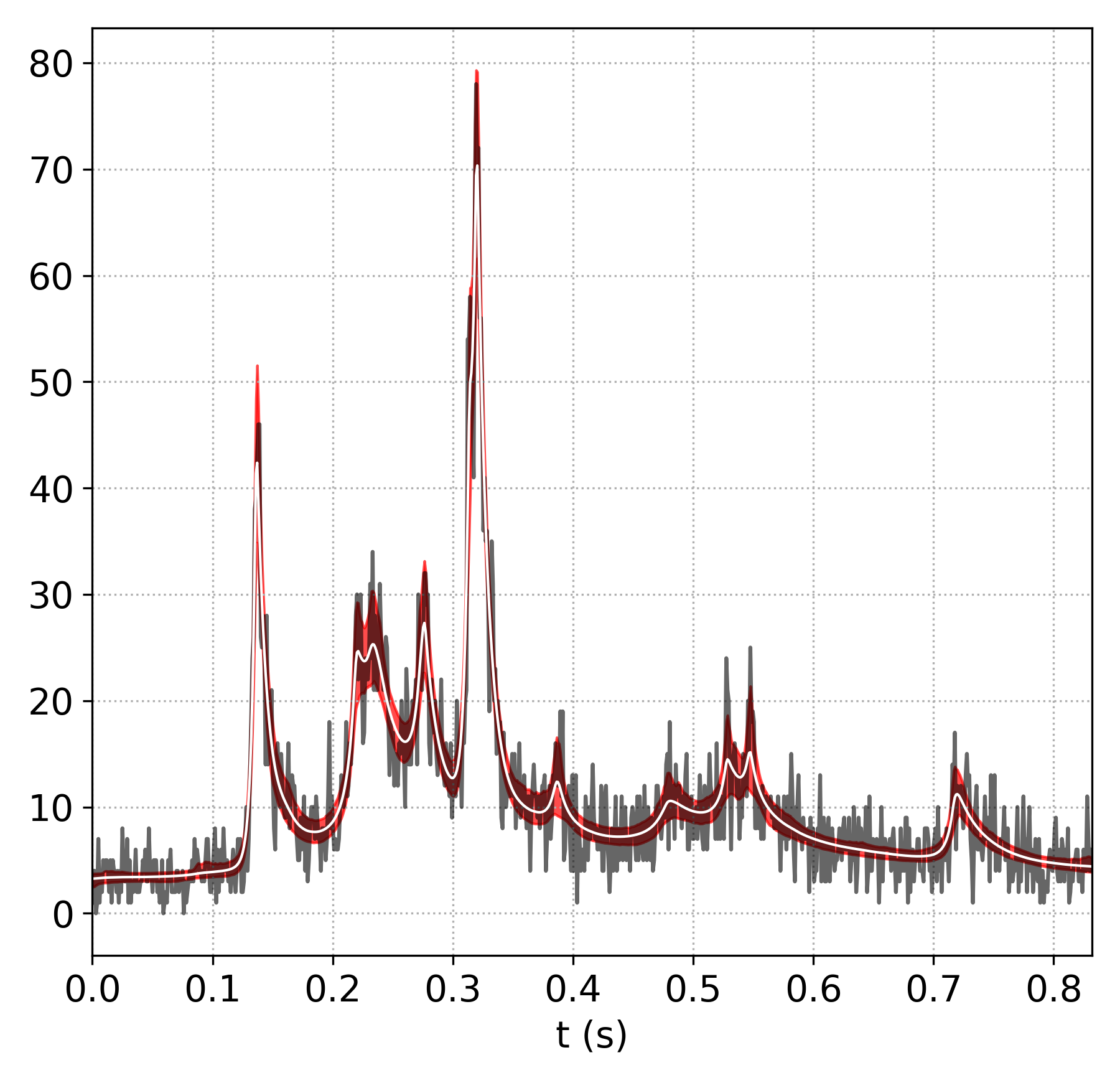}
        \caption{Nested Sampling}
        \label{fig:NS-magnetar-paper-plot}
    \end{subfigure}
    \caption{The mean of 1000 FM and 1006 NS posterior predicted light curves (white) with the 95\% posterior credible interval (red) for an observed burst from magnetar SGR J1550-5418 (trigger ID 090122173). The observed counts were down sampled $\Delta t=5 \cdot 10^{-4}\text{s}  \rightarrow \Delta t=1 \cdot 10^{-3}$s and, for t-FMPE, symmetrically padded with noise to resemble training data.}
    \label{fig:FM-NS-magnetar-curves-comparison}
\end{figure}
\begin{figure}
    \centering
    \begin{subfigure}{0.25\textwidth}
        \centering
        \includegraphics[height=1.5in]{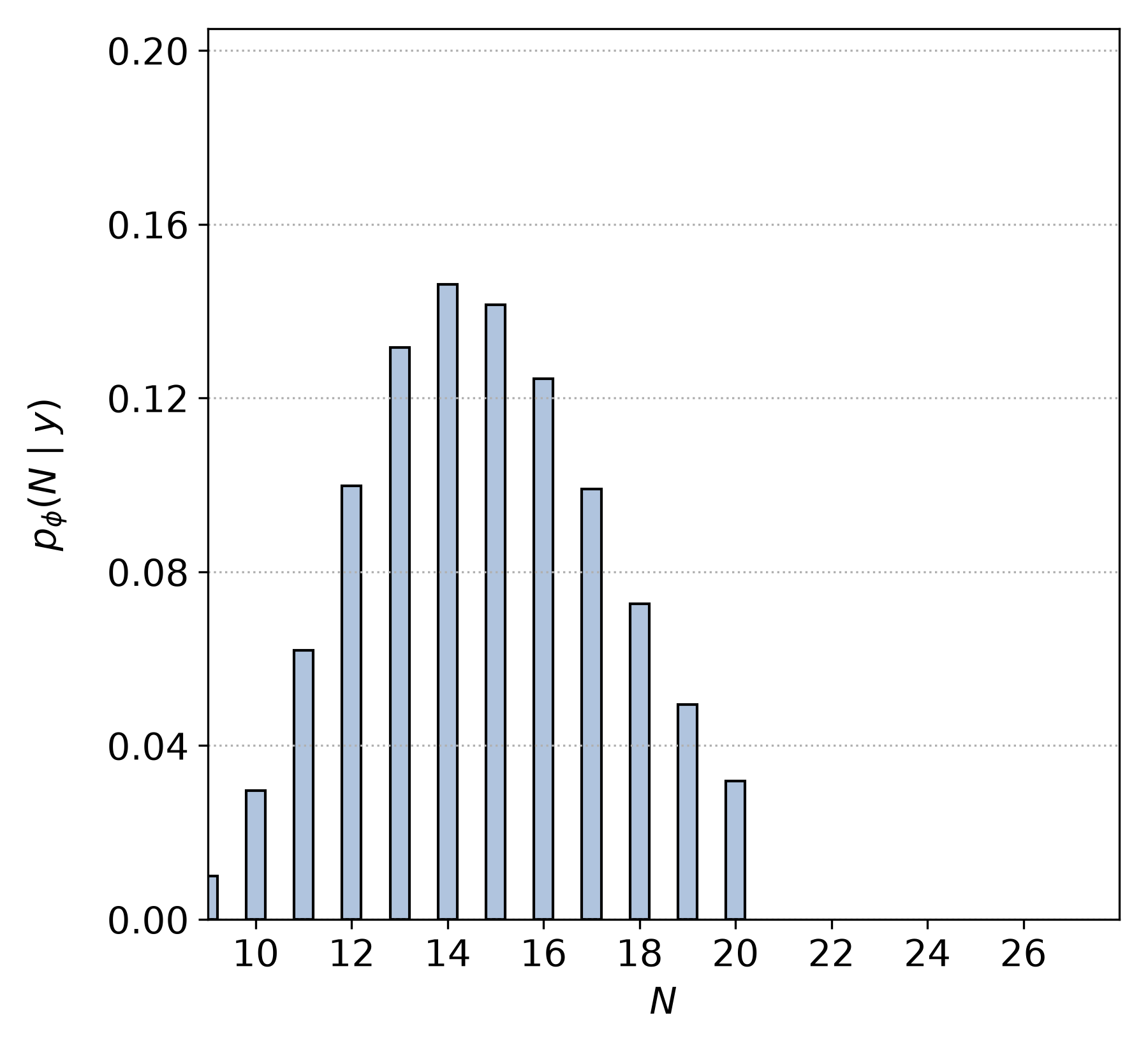}
        \caption{Flow Matching}
        \label{fig:FM-magnetar-paper-plot}
    \end{subfigure}%
    ~
    \begin{subfigure}{0.25\textwidth}
        \centering
        \includegraphics[height=1.52in]{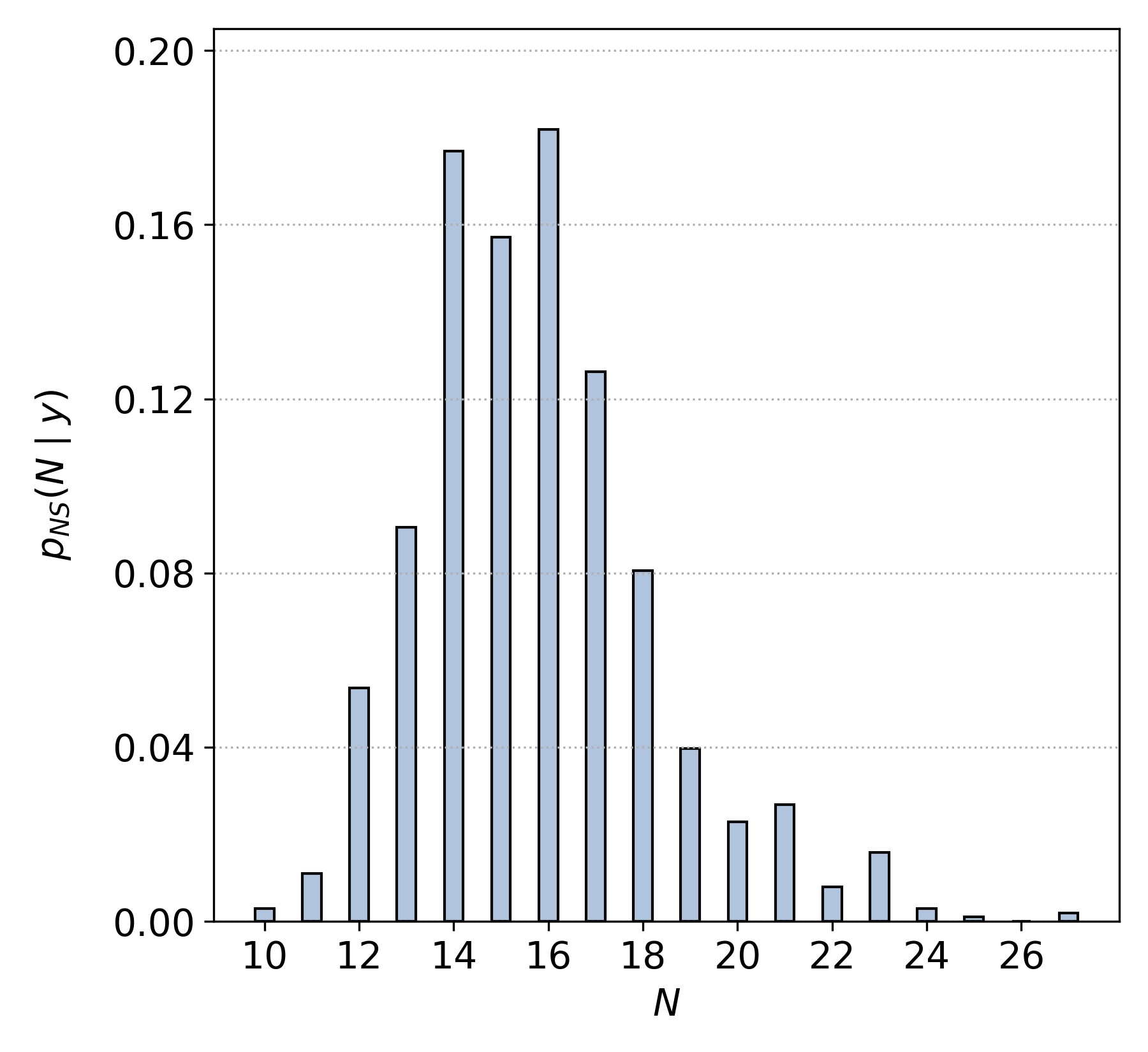}
        \caption{Nested Sampling}
    \label{fig:NS-magnetar-components-paper-plot}
        
    \end{subfigure}
    \caption{FM and NS comparison between the predicted number of components in the magnetar burst from figure \ref{fig:FM-NS-magnetar-curves-comparison}. (A): Predicted FM posterior over the components $p_\phi(N\mid y)$. (B): The NS approximate posterior constructed from 1006 draws using the method from \citet{Huppenkothen_2015}, with adjusted priors to match those used in FM. }
    \label{fig:NS-FM-magnetar-components-comparison}
\end{figure}
We use Network 3 as described in table \ref{tab:network-configurations}, which is trained on bursts with Poisson noise and $N_{\rm max}=20$ components at most. 
On the A100 GPU, FM inference time was 1 second per 100 samples. Nested sampling took 10 minutes for 1006 samples. Figure \ref{fig:NS-magnetar-paper-plot} shows the mean FM and NS posterior predicted light curve and 95\% credible interval overlaid on the original light curve. The mean light curve is defined as $\{\langle \lambda_k \rangle\}$, where
\begin{equation}
    \langle \lambda_k \rangle = \frac{1}{1000}\sum_i^{1000} \lambda_{k, i}
\end{equation}
with $\lambda_{k, i}$ the flux value in bin $k$ of the $i$-th posterior predicted light curve. The 95\% data interval is defined as the area between the 2.5th and 97.5th percentiles, meaning 95\% of the sampled light curves fall within this interval. For both methods, the mean curve lies within the noise of the data, but the nested sampling results show a higher number of distinct peaks and lower variance, particularly in the interval $t=0.35-0.6$. 

Figure \ref{fig:NS-FM-magnetar-components-comparison} compares the predicted marginal posterior over the number of components. For FM, this posterior is directly available through the classifier's output, while for NS it is constructed from 1006 posterior draws. Both distributions peak around $N=14$ and have similar spread and shape, with the NS distribution showing a longer tail from $N=21$ to $N=27$. This tail is absent in the FM distribution, since the network is configured to detect at most $N_{\rm max}=20$ peaks in a light curve. Taking this into account, the two methods appear to show good agreement.

\subsubsection{FRBs}\label{sec: FRB-FMPE-results}
\begin{figure}
    \centering
    \includegraphics[width=\linewidth]{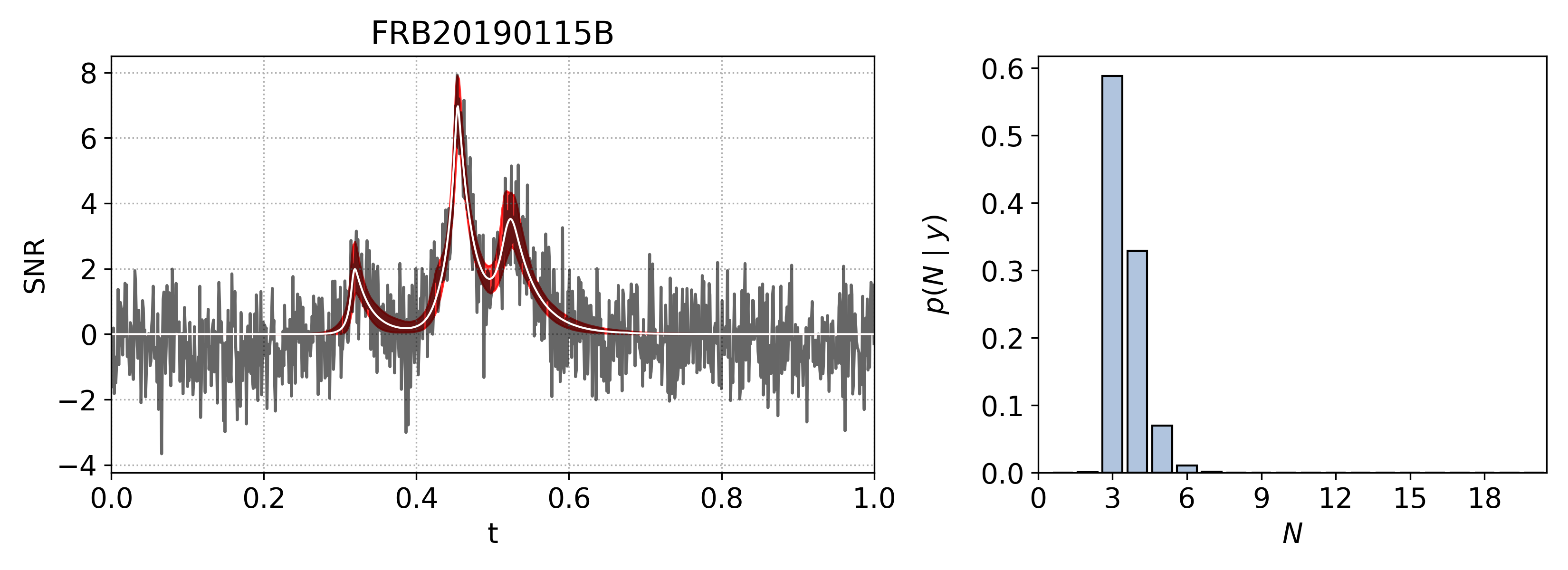}
    \caption{ Left: The mean of 1000 posterior predicted light curves (white) with the 95\% posterior credible interval (red) plotted over pre-processed radio flux of FRB20190115B (black). Right: $p_\phi(N\mid y)$, with $y$ the observed flux. The posterior over the components is well-constrained near $N=3$. }
    \label{fig:FRB20190115B}
\end{figure}

\begin{figure}
    \centering
    \includegraphics[width=\linewidth]{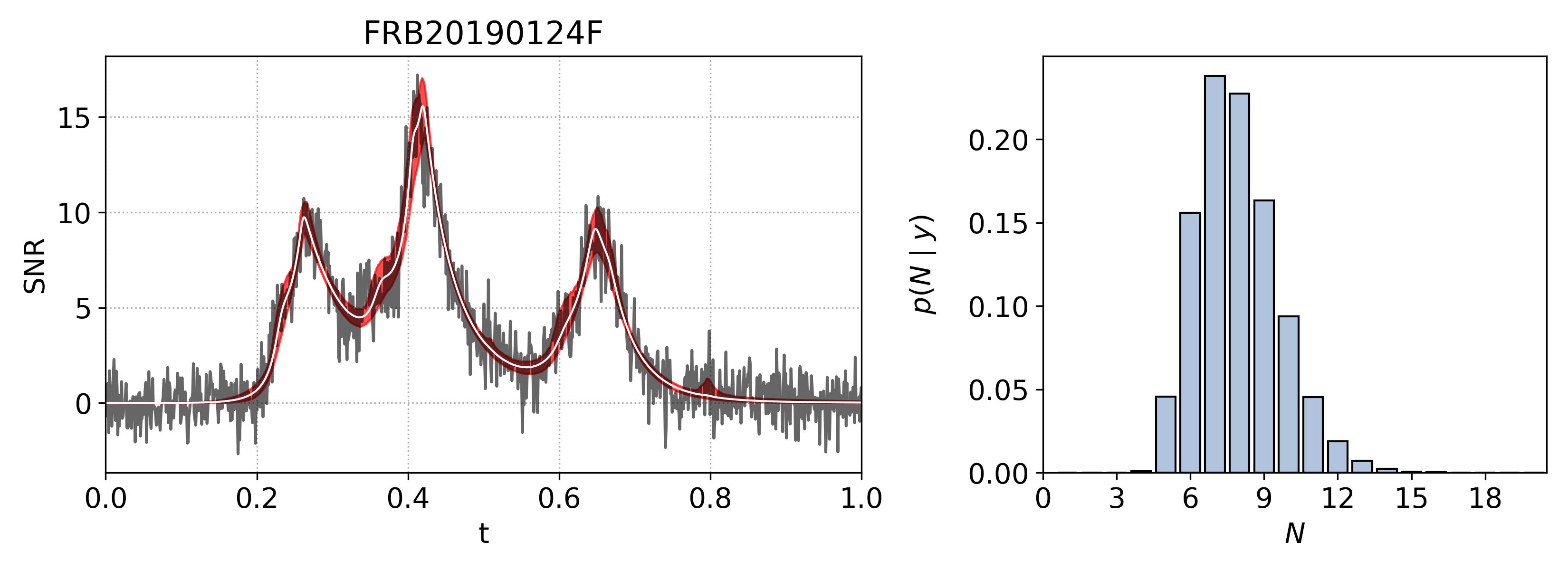}
    \caption{As in figure \ref{fig:FRB20190115B}, but for FRB2019124F. }
    \label{fig:FRB2019124F}
\end{figure}

\begin{figure}
    \centering
    \includegraphics[width=\linewidth]{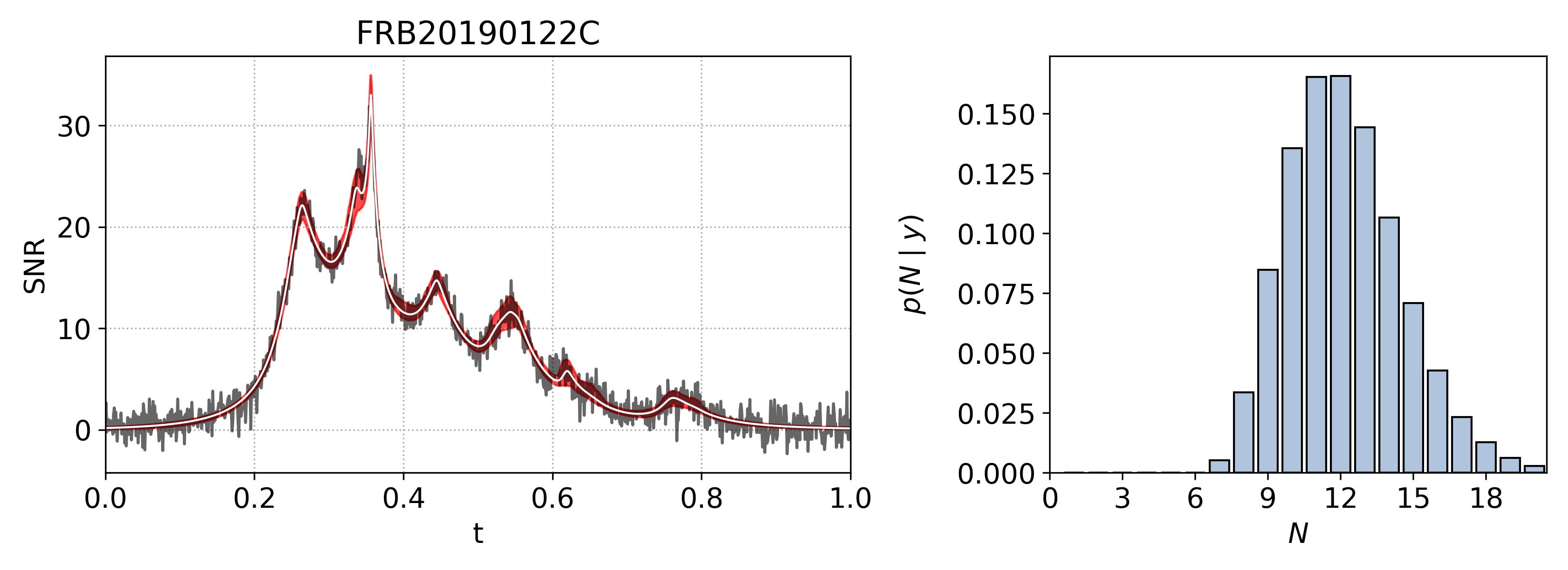}
    \caption{As in figure \ref{fig:FRB20190115B}, but for FRB20190122C. }
    \label{fig:FRB20190122C}
\end{figure}

t-FMPE was applied to the three multi-peaked radio bursts described in section \ref{sec:OD}. Here, we use the same network configuration as in the previous section, now trained on simulated FRB profiles with Gaussian noise and $\lambda_{\rm bkg}=0$ instead (Network 4; see table \ref{tab:network-configurations}). FM sampling speed is again 100 posterior samples per second.
1000 posterior samples were generated from the trained network using algorithm \ref{alg:inference} for the pre-processed flux of FRB2019115B, FRB2019106B and FRB2019122C. Figures \ref{fig:FRB20190115B}, \ref{fig:FRB2019124F} and \ref{fig:FRB20190122C} show two plots for each FRB: The mean posterior predicted light curve with the 95\% posterior credible interval, and the inferred posterior over $N$, $\bm p_\phi(N\mid y)$.  Overall, the predicted light curves are consistent with the observed flux.

\begin{figure*}
    \centering
    \includegraphics[width=0.8\textwidth]{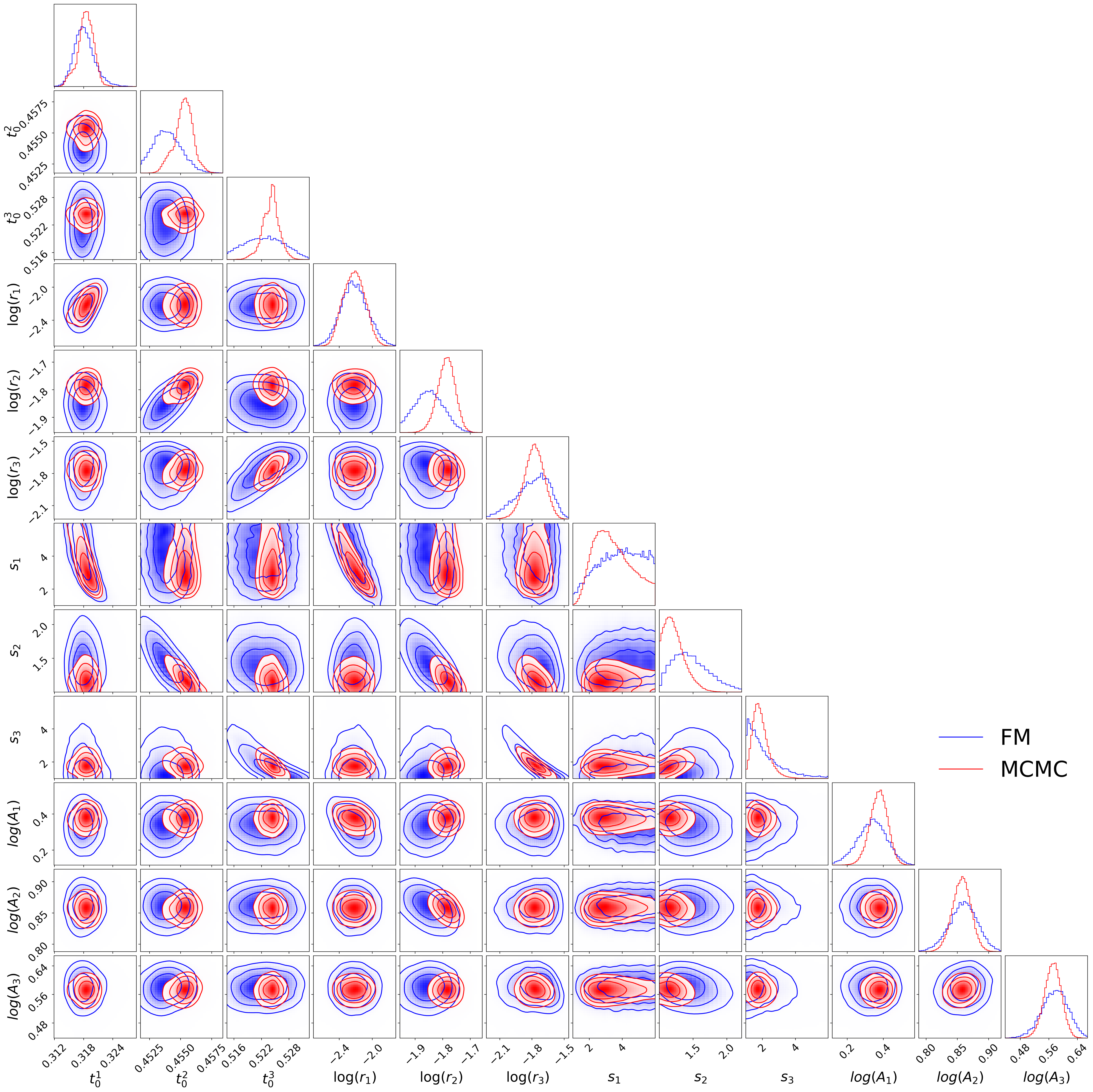}
    \caption{FM (blue) and MCMC (red) posterior distributions for FRB20190115B when enforcing $N=3$. The FM posterior is broader compared to MCMC. The distributions of the peak time and rise time of the second peak ($t_0^2$ and $\log{r_2}$) appear to be slightly off-set.}
    \label{fig:FRB-mcmc-fm-corner-comparison}
\end{figure*}

A reference MCMC posterior was generated for the burst with the lowest and most constrained predicted number of components, FRB20190115B. This choice was made to ensure that the resulting corner plot remains interpretable and the MCMC chains converge within a reasonable time frame. Setting the number of components to $N=3$, the most probable estimate from figure \ref{fig:FRB20190115B}, $9.25\cdot10^5$ MCMC samples were generated using the Gaussian likelihood function. After a burn-in period of 10,500 steps, the sampler reached convergence after 18,500 steps, with a total runtime of 5 minutes. For this same burst, 25,000 FM samples were generated\footnote{We use 25,000 samples to ensure sufficient coverage for a corner plot that clearly shows the shape of the marginal posteriors; with 1000 samples the posteriors are too sparse and noisy.} while enforcing $N=3$, such that $\bm \theta \sim p^\phi(\bm\theta\mid \bm y,N=3)$. A comparison of the resulting posterior distributions is shown in the corner plot in figure \ref{fig:FRB-mcmc-fm-corner-comparison}. The marginal posteriors appear to follow similar trends, with the FM posterior broader than the MCMC posterior, resulting in a C2ST score of 86\%.

The corner plot generated for observed burst FRB2019115B (figure \ref{fig:FRB-mcmc-fm-corner-comparison}) demonstrates a similar difference in posterior spread between MCMC and FM seen in figure \ref{fig:5-peaks-corner-mcmc-fm}. This broadness might indicate a trade-off between the ability of the amortized posterior to generalize across a large and diverse set of light curves and the calibration of any individual posterior. Comparing the corner plots in figures \ref{fig:corner-fixed-1-peak-fm-vs-mcmc} and \ref{fig:5-peaks-corner-mcmc-fm}, which correspond to bursts with $N=1$ and $N=5$ peaks respectively, we find that the posterior overlap between MCMC and FM decreases for the higher-dimensional, five-peaked burst. As both plots were generated using the same network (Network 1; see table \ref{tab:network-configurations}), this could also indicate that higher-dimensional bursts require more training data to get well-constrained posteriors. 

We note that reasonably conservative posteriors are not inherently problematic however, as they are unlikely to exclude true parameter values from their coverage. This is in contrast to \textit{overconfident} posteriors, which can lead to unreliable parameter estimates, and have been noted to emerge in some common SBI algorithms  \citep{hermans2022_a_crisis_in_sim_based_inf}. 

In figure \ref{fig:FRB20190115B} and \ref{fig:FRB2019124F} naive visual inspection would suggest the presence of three components. However, for the burst in figure \ref{fig:FRB2019124F}, the posterior indicates a much larger number of components. This might occur when the burst's shape significantly deviates from the chosen functional form in the component model, in this case a double-sided exponential, requiring a higher amount of components to reconstruct its features. This reconstruction manifests as excessive low-amplitude components in the deviating regions. This effect is especially relevant for observational high SNR bursts where the true component shape is not obscured by noise, such as those observed in \citet{hewitt_2023_dense_forests_microshots_bursts}. We note that the simple component model was deliberately chosen for alignment with previous work in \citet{Huppenkothen_2015}. However, for FRBs in particular, more involved and physics-motivated models exist that may yield better inference results on these bursts \citep{fitburst_algorithm_Fonseca_2024, zhang2023}.

\section{Discussion}

We implemented t-FMPE to solve the challenge of inferring an unknown number of features present in astronomical time series data, which is not solved efficiently through traditional sampling methods. We used a component model that deconstructs light curves into $N$ simple shapes, in this case a double-sided exponential, and aimed to infer the peak time, rise time, skewness and amplitude of each component. t-FMPE was tested on simulated as well as observational data, showing good agreement between posterior samples, data, and reference posteriors generated with MCMC and nested sampling. Here, we discuss current limitations of the model, design choices and future work. 

\subsection{Broadness of t-FMPE posteriors}
As discussed in the previous section, we noticed that the t-FMPE posteriors tend to broaden relative to the reference posteriors as the number of components increases, suggesting that the network struggles more in this regime, although a more thorough evaluation metric (e.g. coverage or simulation-based calibration) would be needed to confirm this. It is not obvious what could be causing this. Possible ways to mitigate this include skewing the prior of the number of components in order to include more multi-peaked bursts in the training data, using better informed priors that include correlations between parameters of different peaks, or a more expressive time series encoder. The current multi-scale CNN uses convolutions that might not be optimal for representing complex multi-peaked light curves with long range correlations between peaks. Again, we note that broad posteriors are not necessarily problematic, but further testing is needed to verify the t-FMPE posteriors are genuinely conservative. 

\subsection{Network Setup}
Currently, the network relies on token masking to manage the trans-dimensional aspect of the problem. However, this approach introduces some computational inefficiency for cases where $N_{\rm true}$ is small compared to $N_{\rm max}$. Although tokens are masked and consequently excluded from self-attention, this does not reduce the overall computational cost of a forward pass. This means that if the number of components in a collection of time series were to follow a heavy-tailed distribution (e.g. a power-law), a large portion of computation during training is effectively wasted on masked tokens. It is not clear how this inefficiency can be addressed during training, as parallelization requires batches with fixed dimensions, in this case $\bm\theta \in {\rm I\!R}^{B\times 4N_{\rm max}}$. One possibility would be to group batches based on the value of $N$, in order to minimize the fraction of masked tokens, though this could introduce additional computational overhead from sorting. 
At inference however, the sampling speed can be increased by setting $N_{\rm max}$ to the largest non-zero value of $\bm p_\phi(N\mid \ y)$. Since the computation time of self-attention scales quadratically with the sequence length $N_{\rm max}$, doing so should speed up inference significantly, especially for less complex, low component bursts. 

We also note the classifier and transformer are currently optimized independently, such that the transformer part of the network is unable to affect the number of components it has been given. We argue that this is not necessarily problematic, as uncertainty in $N$ is propagated into the parameter estimation stage during inference, instead of collapsing the outcome of the classifier onto a single value. 

Furthermore, the two conditions, light curve $\bm y$ and flow matching time $\tau$ are currently concatenated into each token. This approach is straightforward and has demonstrated good results within the current setup. While it does introduce some redundancy, since the same information is repeated across all tokens, it has been effective in practice and provides a simple way to incorporate conditions without increasing the architectures' complexity. This same method has been used for classification of astronomical transients to explicitly include additional informative features of the signal, such as the redshift \citep{allam2023_paying_attention_astronomical_transients}. 

\subsection{Consequences of chosen priors}
In this pilot study, several choices were made to simplify the learning task, such as restricting the peak times in the training data to be between 0.2 and 0.8 seconds and fixing the length of the time series to 1000 bins. Because of this, the real observational time series had to undergo minor adjustments in order to accommodate these choices. In practice however, it is advisable to first gain insight into the dataset of bursts to be studied, and define relevant prior distributions accordingly, ensuring the training data matches the observational data such that it can be used without adjustments. 

Ideally, observational data is supported by the parameter ranges of the selected priors, for example by having amplitudes between 1 and 300 in this case. The network does not generalize to data outside of the priors used during training, and should be re-trained with adjusted priors when such mismatches are identified. An example of a potential mismatch is illustrated by the marginal FM posteriors of $s_{2,3}$ in figure \ref{fig:FRB-mcmc-fm-corner-comparison}. Here, the posterior mass is pushing up against the lower bound of the skewness prior, suggesting that the selected prior range may be too narrow. This lower bound of 1 on skewness was originally chosen under the assumption that the rise time should be smaller than or equal to the fall time due to scattering effects; however, this result indicates that this assumption may need to be reconsidered. 


\subsection{Sampling Speed}
t-FMPE outperforms both t-NS and MCMC in computational efficiency when sampling high-dimensional space. Directly comparing MCMC and t-FMPE inference times is not straightforward; MCMC sampling requires a burn-in period whose duration can not be predicted in advance and can vary between runs. Generally, for low-dimensional cases, the overall inference time is comparable between the two methods. However, as dimensionality increases, MCMC performance slows down significantly, while t-FMPE inference time remains effectively constant. For \texttt{magnetron}, which uses nested sampling, the comparison is more direct; nested sampling was continued until $\sim$1000 samples were reached, which required 10 minutes. With FMPE, this same number of samples is generated within 10 seconds on an A100 GPU, reaching a sampling speed that is approximately 60 times faster than the \texttt{magnetron} sampling algorithm. 

This is all without considering the upfront training cost of the FM model, which is about a day for the largest network used. Training time mostly depends on the size of the network and the amount of training data required for reasonable convergence of the loss, which typically increases with the maximum number of burst components $N_{\rm max}$. When considering the training cost, there are two cases where traditional sampling methods could be preferable in terms of computational cost, under the condition that the posterior can be evaluated: Low-dimensional posteriors and cases where only a single approximate posterior is needed. But when one needs posterior estimates for a large dataset of bursts, t-FMPE is sure to be the more efficient than MCMC or t-NS.  

\subsection{Processing of Observational Data}
\label{sec:obsprocessing}
Post hoc adjustments of observational data in this study include downsampling and padding with noise, before passing through the network. This was done to ensure a match with the training data in terms of number of bins and peak location, which is set to be between 0.2 $\leq t_0 \leq$ 0.8 seconds. No training data is generated with components outside of these bounds, and hence any out-of-bounds peaks will go undetected by the network. These bounds were originally chosen for two reasons. First, to simplify the training task by avoiding component features that lie largely out of range. Second, to simulate the way observational bursts are typically cut out from a much longer time series and centred in a time window surrounded by some background noise. However, this approach does not account for cases where pre-processing algorithms might miss especially faint burst components that extend outside of these bounds. To reduce the need for post hoc padding, the peak time bounds could be extended without issue. 

The other reason for padding and downsampling was to match the training data in terms of number of bins $K$. This is necessary because the time series encoder uses convolutions and average pooling, which are not invariant to the input length. Hence attempting inference on light curves with a different number of bins can lead to poor agreement between the model and the data. This can be problematic for exceedingly short or long observed bursts, which could not be reliably analysed with the current t-FMPE model. 

A final element in our currently trained networks limiting generalization to observational data, is how the background rate $\lambda_{\rm bkg}$ must be specified in advance. A mismatch between its value in the training and test data will lead to inaccurate inference. This issue can be mitigated by requiring all inputs to be background-subtracted. 

\subsection{Future work}\label{sec:future-work}
In addition to the suggestions made above, we note that the existing setup can be meaningfully expanded on two fronts; the burst model and the evaluation metrics. 

\subsubsection{The Burst Model}
First, the current burst model uses a double-sided exponential as the functional form of individual components, for simplicity and for straightforward comparison with prior work. It could be beneficial to experiment with a set of functional forms instead, for example a skewed Gaussian, which is commonly used to fit FRBs in literature. For high SNR bursts, which have been observed as structures seemingly made up of a set of distinct shapes \cite{hewitt_2023_dense_forests_microshots_bursts}, exponential components might fail to effectively capture the burst shape, leading to an excess of low amplitude components. Using a mixture of shapes could potentially mitigate this issue. 

Furthermore, t-FMPE, being likelihood-free, permits more realistic simulations. Effects that would make traditional methods intractable, can now be included in the simulator, such as physical and instrumental effects, allowing for more thorough testing of the underlying physics. For instance, physics injections could focus on scattering effects in the tails of burst components. An instrumental effect to be considered is dead time, a phenomenon in X-ray and $\gamma$-ray astronomy where the observed number of counts is lower than the true value due to over-saturation of the detector \citep{dead_time_Bachetti_2015}. Similarly, in radio, uncertainties in the DM of an FRB are often not propagated into the burst modelling stage, potentially yielding biased burst parameters. Incorporating DM into the simulator could effectively mitigate that problem.

Something else to be considered in the case of FRBs is that any frequency-dependent information encoded in the full dynamic spectrum is currently being neglected by using the integrated flux. However, the dynamic spectrum holds valuable information, it has for instance been noted that the components of multi-peaked bursts tend to be concentrated in different frequency bands \citep{chime_second_catalogue_2023}. Therefore, it is useful to explore the option of replacing the integrated flux $\bm y$ with the full dynamic spectrum $D(\nu, t)$. There fortunately already exists a computational model for $D(\nu, t)$, as part of the `\texttt{fitburst}' algorithm \citep{fitburst_algorithm_Fonseca_2024}. The free model parameters of `fitburst' also include global parameters such as scattering time scale and dispersion measure, which are not accessible when only using the integrated flux. Therefore, including the full dynamic spectrum is expected to enable inference of more physically informative parameters. To facilitate this, the one-dimensional convolutions in the time series encoder could trivially be replaced by two-dimensional convolutions. It should be noted that adding an extra dimension to the training data will significantly increase computational overhead, making the on-the-fly generation of training data likely no longer possible. 

A significant limitation of the current implementation is the requirement that the number of bins in flux must be $K=1000$, whereas in practice this will depend on the duration of each detected burst. Both magnetar bursts and FRBs occur on a range of timescales, making it challenging to work with a fixed number of bins. This can potentially be partially resolved by training on inputs of variable length, since the time series encoder can handle variable input via adaptive average pooling. However, the encoder is not expected to handle substantial differences in length (e.g. $> 20$\%), though this tolerance remains to be evaluated in future work. If bursts do vary considerably in length, it may be beneficial to employ a different architecture for the time series encoder to accommodate this. Exceptionally long bursts could be segmented into shorter sections before processing. 

\subsubsection{Evaluation Metric} 
At present, there is no trivial way to evaluate the performance of the full amortized trans-dimensional posterior generated with t-FMPE. In this work, evaluation was based on a small subset of bursts with available reference posteriors, which was sufficient to demonstrate proof-of-concept of the proposed method. However, in practice, it is essential to assess network performance across a representative part of the parameter space. Such a metric is needed to compare network architectures, conducting hyperparameter sweeps, and ensuring generalization to a variety of bursts. 

However, developing such a performance metric is challenging. Many existing methods rely on direct evaluation of posterior probabilities or become computationally prohibitive for high-dimensional samples. Approaches have been proposed that avoid posterior evaluation and only require posterior samples, such as the C2ST used in this work \citep{lueckmann-2021-benchmarking-sbi}. While C2ST is relatively simple to implement and interpretable, in its current form it is not ideal for this setting; it still requires access to reference posteriors, and only evaluates one example from the amortized posterior. This may be avoided however by modifying the C2ST to compare joint samples ($\bm \theta, \bm y$) of the forward model (the simulator) to those of the reverse model (the neural posterior) instead. This could enable quantifying the performance of the amortized posterior across large set of light curves, without the need for a reference posterior. 

Another option to explore in future work is simulation-based calibration (SBC), which is one of the few approaches that does not require a reference posterior or posterior evaluations, while also incorporating an ensemble of simulations \citep{talts-2020-sbc-validatingbayesianinferencealgorithms}. Although SBC was excluded from the SBI benchmark \citep{lueckmann-2021-benchmarking-sbi} due to its requirement of repeated inference on many observations, which made it prohibitively slow for most algorithms, this should not apply to t-FMPE. With SBC, the calibration of one-dimensional posteriors is checked through a histogram of the rank of the true parameter value within the posterior samples, estimated over many simulations. For uniform priors, deviations from a uniform histogram indicate poor calibration: a U-shaped histogram indicates an overconfident posterior, while parabolic shapes suggest a conservative posterior. Exploring SBC and other metrics can improve the evaluation of amortized t-FMPE posteriors in the future, allowing for reliable hyperparameter sweeps and statements on generalization capability of the network. 


\section{Conclusions}

This work introduced the challenge of sampling a trans-dimensional posterior distribution of fast transient light curves, where the number of burst components is not known in advance. For FRBs especially, fast and accurate inference is increasingly important with the amount of observational data available rapidly increasing in recent years. This dataset is expected to continue growing exponentially, lending itself to data-driven discovery of FRB progenitors and emission models. However, traditional statistical methods often require explicit expressions for the likelihood, rely on manual tuning and summary statistics and suffer from the curse of dimensionality, leading to intractable, computationally expensive or very involved sampling. 

We therefore presented trans-dimensional Flow Matching Posterior Estimation for fast amortized inference of burst parameters. In this method, the posterior is approximated by using simulated data to learn a time-dependent vector field that, guided by an observation, induces a flow from the Gaussian base distribution into the posterior. 

To facilitate the trans-dimensional aspect, t-FMPE was implemented on a scalable transformer architecture, where the effective sequence length of the input is estimated by a separate module referred to as the classifier. This is uniquely possible in t-FMPE because, contrary to other neural posterior estimation methods, no restrictions are placed on the neural architectures used. 

Results showed good agreement with both simulated and observed flux through posterior predicted profiles. For given $N$, t-FMPE posterior distributions demonstrated substantial overlap with MCMC reference posteriors, and accurately captured observed relationships between parameter pairs, such as the linear relation between rise time and skewness. t-FMPE posteriors appeared generally conservative, likely as a result of the need for generalization capability across a large and diverse set of light curves or lack of training data, although further testing is required to verify this. 

When comparing to \texttt{magnetron}, a trans-dimensional nested sampling method developed for magnetar bursts by \citet{Huppenkothen_2015}, t-FMPE produced qualitatively similar results in a fraction of the time, requiring only 10 seconds for 1000 posterior draws instead of 10 minutes. This method therefore shows potential to greatly simplify and accelerate the analysis of large volumes of time series data, lowering the threshold to in-depth studies of temporal variability. We also note that t-FMPE is not necessarily restricted to time series data and might be modified to suit any trans-dimensional inference problem that deals with sequential data.
\section*{Acknowledgements}


The authors thank \href{www.surf.nl}{SURF} for the support in using the Dutch National Supercomputer Snellius as a computational resource. We thank the CHIME/FRB collaboration for publicly releasing the fast radio burst data used in this work.

\section*{Conflict of Interest}


The authors declare no conflict of interest.
\section*{Data Availability}
The source code is publicly available \href{https://github.com/Ninavd/FRB-flow}{here}, including the pre-processed FRBs and magnetar burst. The unprocessed FRB files are available on \href{https://www.canfar.net/storage/vault/list/AstroDataCitationDOI/CISTI.CANFAR/23.0029/data/beamformed_files}{Canfar}. 




\bibliographystyle{rasti}
\bibliography{References} 




\appendix


\section{Implementation Details} 

\label{AppendixA} 
\label{sec:implementation-details} 
This section focuses on implementation details of the t-FMPE method required for reproduction of this study, but not essential for conceptual understanding of the method. 

\subsection{Hyperparameter overview}
Table \ref{tab:hyperparams} provides an overview of hyperparameters used in training and their values. 

\begin{table}
\centering
\begin{threeparttable}
\caption{Hyperparameter values.}
\label{tab:hyperparams}
\begin{tabular}{lll}
\toprule
 Symbol &Value& Description\\ 
 \midrule
 $B$&1024& Batch size\\
 $n_{\rm steps}$& 200&Number of integration steps at inference\\
 $\alpha$& $4/3$&Exponent in sampling distribution of $\tau$ \\
 $\sigma_{\rm min}$& $10^{-4}$&Minimum variance of Gaussian path at $\tau=1$\\
 clip& 1 &Clipping value of gradient norms\\
 \bottomrule
\end{tabular}

\end{threeparttable}

\end{table}

\subsection{Time Series Embedding Network}
Here, we discuss in detail the embedding networks used to compress time series $\bm{y}$, before being processed into tokens. 

\begin{figure*}
    \centering
    \includegraphics[width=0.8\linewidth]{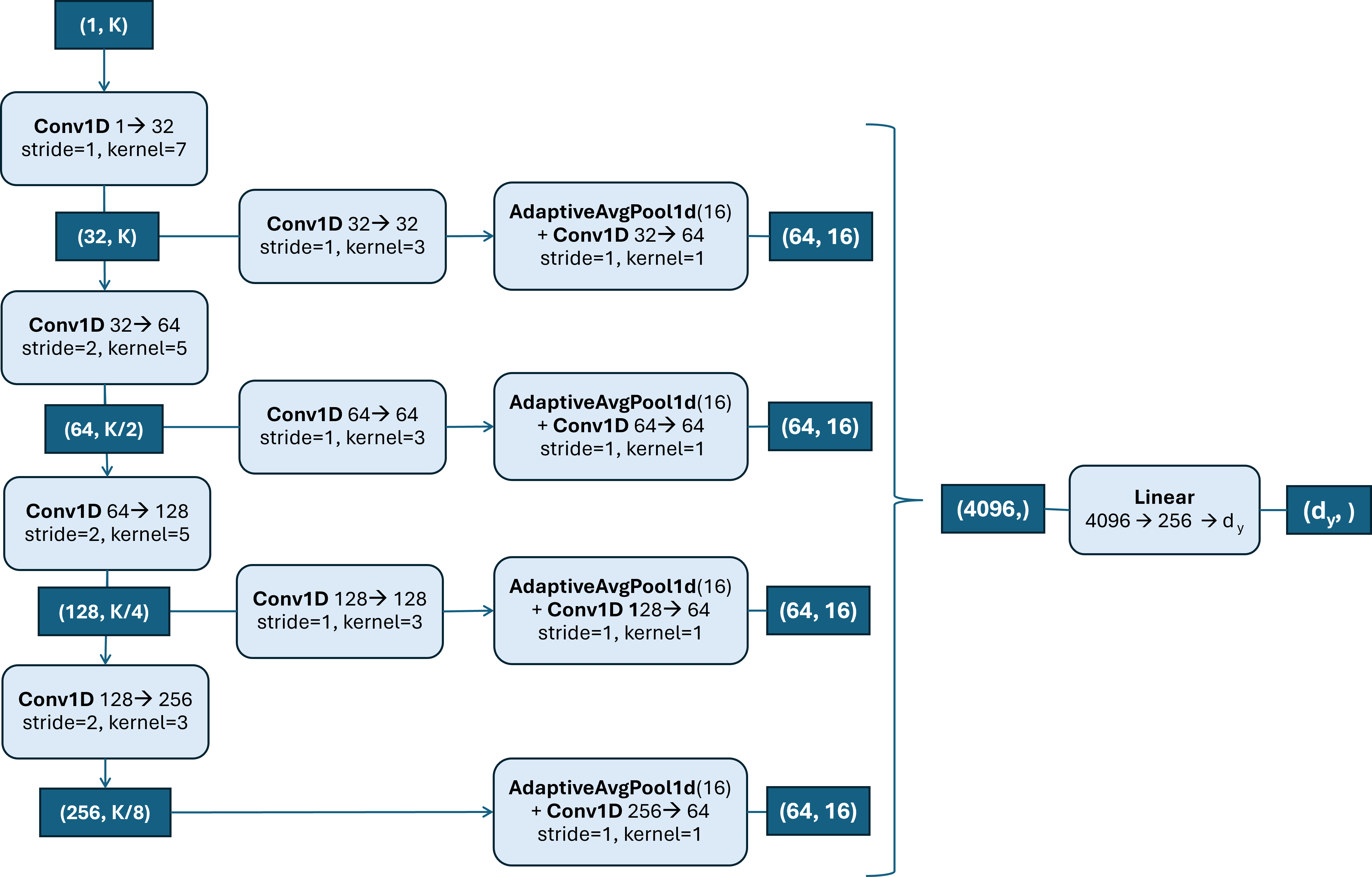}
    \caption{Multi-scale CNN encoder architecture used to encode the burst profile. The output dimension of each block is indicated in the dark-blue rectangles as (\#channels, length). The input is the burst profile $y \in R^K$, indicated in the top rectangle. The latent dimension is $d_y \in [64, 128]$, depending on the settings.}
    \label{fig:detailed-UNET}
\end{figure*}

The CNN encoder maps the burst time series of length $K$ into a latent vector of dimension $d_y$. The input passes through four successive one-dimensional convolutional layers (see figure \ref{fig:detailed-UNET}):
\begin{enumerate}
    \item kernel size 7, stride 1, 1 $\rightarrow$ 32 channels
    \item kernel size 5, stride 2, $32 \rightarrow 64$ channels
    \item kernel size 5, stride 2, $64 \rightarrow 128$ channels
    \item kernels size 3, stride 2, 128 $\rightarrow$ 256 channels
\end{enumerate}
where ReLU is applied after each layer. To the output of layer 1-3, an additional one-dimensional convolution is applied while preserving length and channel dimension. 

Adaptive average pooling is applied to the outputs of each layer, to get a vector of fixed length of 16 in each channel. This is followed by a $1\times 1$ convolution to project to a fixed number of channels, 64. This gives representations of each scale with the same number of channels (64) and length (16). These feature vectors are flattened to (B, 1024) and concatenated into (B, 4096) before passing through two linear layers projecting from (B, 4096) to (B, 256) to (B, $d_y$) with ReLU activation, where $d_y$ is the user-specified latent dimension. 

\subsection{Classifier}
Recall that the classifier consists of a multi-scale CNN encoder, followed by an MLP, as was illustrated in figure \ref{fig:classifier-diagram}. The MLP has layers of sizes [$d_y$, $d_y$/2, $d_y$/2, $d_y$/4, $d_y$/4, $N_{\rm max}$], which use the SiLU activation function, except for the output layer, which uses Softmax to generate a normalized probability vector. 
The architecture of the CNN encoder is described in the section above. If training data is configured to contain only fixed number of components, the classifier is removed from the model, and $N_{\rm true}=N_{\rm max}$ always holds.

\subsection{Transformer Encoder}
\begin{figure}
    \centering
    \includegraphics[width=0.3\linewidth]{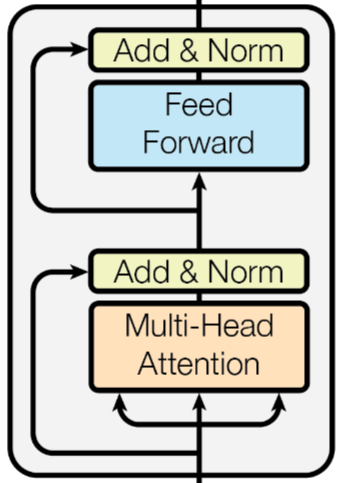}
    \caption{Structure of a single encoder block. One block consists of multi-headed self-attention and a feed forward layer with residual connections and layer normalization. Figure adapted from \citet{vaswani2023attentionneed}.}
    \label{fig:encoder-block}
\end{figure}

The transformer encoder described in section \ref{sec:transformer-encoder-method-small} is built using PyTorch's \texttt{Transformer Encoder} and \texttt{TransformerEncoderLayer}, which implements the transformer architecture from the original `Attention is All You Need' paper \cite{vaswani2023attentionneed}. The encoder consists of $L$ sequential encoder blocks. The structure of a single block is shown in figure \ref{fig:encoder-block}. As training data is generated on-the-fly, the dropout rate is set to zero. The dimension of the feed forward layer is set to 1024. We use two different network sizes, depending on the pre-specified value of $N_{\rm max}$. For $N_{\rm max} < 10$, the number of encoder blocks $L$ and attention heads $M$ are 6 and 4 respectively, with a token dimension of 128. For $N_{\rm max} \geq 10$, $L=12$, $M=8$, and $d_t=256$. Irrelevant tokens $\{T_i\}_{i>4N}$ are masked out from attention via the \texttt{src\_key\_padding\_mask} argument, by passing the inverse of the mask in figure \ref{fig:token-mask}. 

\subsection{Scaling}
Burst profiles are scaled down for training stability by the average amplitude of the training data $\bm y' = \bm y /\bar{A}$, with $\bar{A}=150$. 
Targets $\bm \theta_1$ are standardized using 
\begin{equation}
    \bm\theta_1' = \frac{\bm\theta_1 - \bm\mu}{\bm\sigma}
\end{equation}
with $\bm \mu,\bm \sigma \in R^{N_{\rm max}}$ the estimated mean and standard deviation of $\theta_1$, such that the transformed elements have a mean of zero and standard deviation of one. To revert back to the original units at inference, the inverse of this formula is applied to the result of integration. As training data is generated in place, the mean and standard deviation are estimated from 10,000 prior samples before training.  

\subsection{Learning rate}
\begin{figure}
    \centering
    \includegraphics[width=0.7\linewidth]{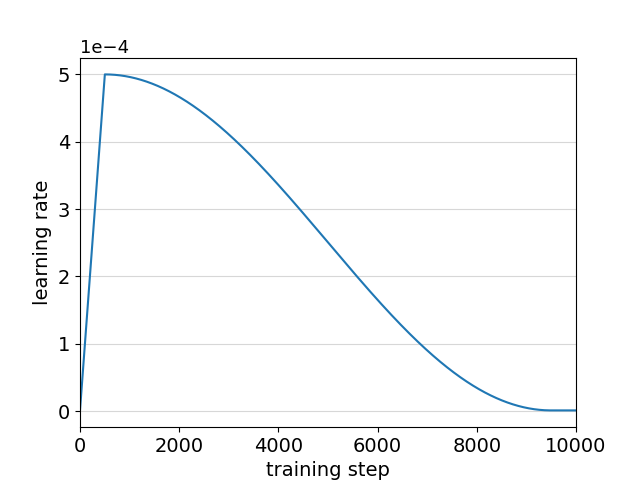}
    \caption{Learning rate value as training progresses. The learning rate increases linearly for the first 500 steps from $10^{-8}$ to $5\cdot10^{-4}$, then decays to $10^{-6}$ through cosine annealing. Example for 10,000 training steps.}
    \label{fig:LR-schedule}
\end{figure}

The learning rate schedule is shown in figure \ref{fig:LR-schedule}. In the first 500 steps, the learning rate is increased linearly (warm up) from 1e-8 to 5e-4 for training stability, after which a Cosine Annealing schedule is used to gradually lower it to 1e-6. The learning rate is fixed for the final 500 steps.

\section{Additional Results}

\begin{figure*}
    \centering
    \includegraphics[width=\linewidth]{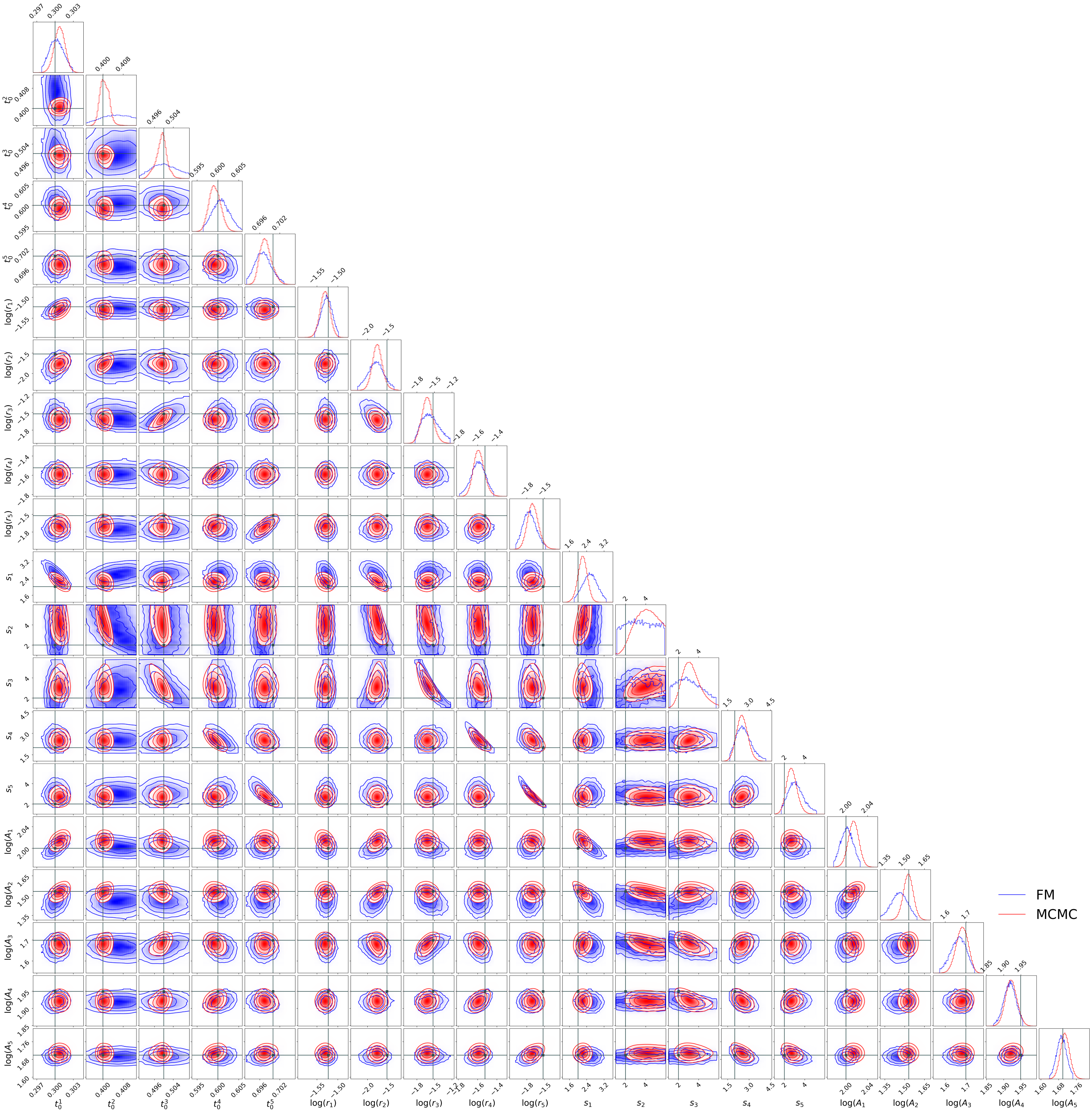}
    \caption{Corner plot comparison of MCMC and FM posterior distributions for the simulated burst in figure \ref{fig:samples-5-peaks-mcmc-fm}.}
    \label{fig:5-peaks-corner-mcmc-fm}
\end{figure*}












\bsp	
\label{lastpage}
\end{document}